\documentclass{aastex631}
\usepackage{url} 
\usepackage{amssymb}
\usepackage{amsfonts}
\usepackage{amsmath}
\usepackage{graphicx}

\newcommand{\eq}{\begin{equation}}
\newcommand{\qe}{\end{equation}}
\renewcommand{\bar}[1]{\overline{#1}}

\newcommand{\e}[1]{\times 10^{#1}}
\newcommand{\B}{\mathcal{B}}
\newcommand{\uvec}{\mathbf{v}}
\newcommand{\heavi}{\mathcal{H}}
\newcommand{\ab}[1]{\langle{#1}\rangle}

\begin{document}

%
%


\title{Energetic constraints on ocean circulations of icy ocean worlds}

%
%

\author{Malte F. Jansen}
\affiliation{Department of the Geophysical Sciences, The University of Chicago, Chicago, IL}
\author{Wanying Kang}
\affiliation{Department of Earth, Atmospheric and Planetary Sciences, Massachusetts Institute of Technology, Cambridge, MA}
\author{Edwin Kite}
\affiliation{Department of the Geophysical Sciences, The University of Chicago, Chicago, IL}
\author{Yaoxuan Zeng}
\affiliation{Department of the Geophysical Sciences, The University of Chicago, Chicago, IL}







\correspondingauthor{Malte F. Jansen, Department of the Geophysical Sciences, The University of Chicago, Chicago, IL}



%

\begin{abstract}
Globally ice-covered oceans have been found on multiple moons in the solar system and may also have been a feature of Earth's past. However, relatively little is understood about the dynamics of these ice-covered oceans, which affect not only the physical environment but also any potential life and its detectability. A number of studies have simulated the circulation of icy-world oceans, but have come to seemingly widely different conclusions. To better understand and narrow down these diverging results, we discuss energetic constraints for the circulation on ice-covered oceans, focusing in particular on Snowball Earth, Europa, and Enceladus. Energy input that can drive ocean circulation on ice-covered bodies can be associated with heat and salt fluxes at the boundaries as well as ocean tides and librations. We show that heating from the solid core balanced by heat loss through the ice sheet can drive an ocean circulation, but the resulting flows would be relatively weak and strongly affected by rotation. Salt fluxes associated with freezing and melting at the ice sheet boundary are unlikely to energetically drive a circulation, although they can shape the large-scale circulation when combined with turbulent mixing. Ocean tides and librations may provide an energy source for such turbulence, but the magnitude of this energy source remains highly uncertain for the icy moons, which poses a major obstacle to predicting the ocean dynamics of icy worlds and remains as an important topic for future research.
\end{abstract}

%
%

%


%
%
%
%

\section{Introduction}

Globally ice-covered oceans have been found on multiple moons in the solar system \citep[][]{Carr1998,Kivelson2000,Thomas2016,Nimmo2016} and spark our curiosity in part due to their potential to provide hospitable environments for life \citep[][]{DesMarais2008,Waite2017,Postberg2018,Hendrix2019}. Earth's oceans may also have been covered by a global ice sheet during the so-called ``Snowball Earth'' events, and indeed eukaryotic life not only appears to have survived through these episodes, but may have evolved significantly during them \citep[e.g.][]{Hoffman2017}. However, relatively little is known about these oceans beyond their existence, and due to our inability to directly observe them at present, we heavily rely on models to decipher their mysteries. 

Although their potential biology may represent the holy grail for research on the icy moon oceans, it is natural to start with the somewhat better constrained problem of inferring the physical and chemical environment. In this study we specifically focus on the ocean circulation and mixing processes, which control the transport of heat and chemical tracers, including those that may affect life and our ability to observe its signatures [e.g. via material ejected in plumes]. 

Ocean circulation on icy moons can broadly speaking be driven by heat and salt fluxes, tidal forcing, or magnetic forces \citep[e.g.][and references therein]{Soderlund2020}. We here focus primarily on ``buoyancy driven" flow, i.e. flows associated with temperature and salinity gradients, although we also include a discussion of the role of tides and librations in driving vertical mixing, which in turn affects the buoyancy field and associated flow \citep[e.g.][]{WunschFerrari2004}. Following most of the previous work on buoyancy-driven flows, we will neglect magnetic forces, although they may be significant on Jupiter's moons \citep{Gissinger2019}.

A number of studies have simulated the buoyancy-driven dynamics of ice covered oceans both in the context of Snowball Earth \citep[e.g.][]{Ashkenazy2013,AshkenazyTziperman2016,Jansen2016} and icy moons \citep[e.g.][]{Soderlund2014,Soderlund2019,AshkenazyTziperman2020,Kang2020,ZengJansen2021,Kang2021}, and seem to have come to widely different conclusions, in particular with regards to the characteristic current speeds in these oceans.  The Snowball Earth simulations of  \cite{Ashkenazy2013}, \cite{AshkenazyTziperman2016} and \cite{Jansen2016} broadly consistently show small Rossby-number turbulent flows dominated by eddies and jets with characteristic velocities on the order of $1\,$cm/s. For Europa, \cite{Soderlund2014} suggest moderate Rossby-number convective turbulence.  \cite{Soderlund2014} report results from their direct numerical simulations (DNS) in terms of non-dimensional velocities (which amount to flow Rossby numbers) and find $|U/(2\Omega D)|\sim 0.5$. Taking these DNS results at face value and re-dimensionalizing with Europa's rotation rate and ocean depth\footnote{Using the rotation rate $\Omega\sim 2\e{-5}\,$s$^{-1}$ and ocean depth $D\sim 1\e{5}\,$m, $|U/(2\Omega D)|\sim 0.5$ translates to $|U|\sim 2.5\,$m/s.} would suggest flow speeds in excess of $1\,$m/s. Using a global circulation model for Europa's ocean, \cite{AshkenazyTziperman2016} instead find largely geostrophic (i.e. low Rossby-number) turbulence and jets with characteristic velocities on the order of $1\,$cm/s. In a parameter regime deemed applicable to Enceladus, DNS by \cite{Soderlund2019} would suggest $|U|\sim 0.1\,$m/s if the non-dimensional velocities are taken at face value\footnote{\cite{Soderlund2019} report non-dimensionalized velocities $|U/(\Omega D)|\sim 0.1$ for an Enceladus-like parameter regime. Rescaling with the rotation rate $\Omega\sim 5\e{-5}\,$s$^{-1}$ and an ocean depth $D\sim 2\e{4}\,$m would yield $|U|\sim 0.1\,$m/s }. Instead \cite{Kang2020} and \cite{ZengJansen2021} find velocities on the order of $0.1\,$mm/s using global-scale ocean simulations. The  flow dynamics and associated kinetic energy levels hence appear to vary widely across these studies, with variations across different studies being larger than variations between different oceans.

To shed some light on these apparent discrepancies and to establish what insights can be gained from first principles (i.e. without running numerical simulations whose results are sensitive to many parameters and assumptions), we here consider energetic constraints for the circulation of a globally ice-covered ocean. For simplicity we limit ourselves to an ocean in a statistical equilibrium state (as also assumed in all studies discussed in the previous paragraph). Specifically, we assume that the net global ocean warming or cooling is small compared to the heat fluxes through the lower and upper ocean boundaries, and similarly that the net global mean freezing or melting rate is small compared to the regional rates. Non-equilibrium effects could be important \citep[e.g.][]{Hussmann2004,Nakajima2019}, but would vastly widen the range of possible solutions. Our philosophy is that the better-constrained equilibrium problem should serve as a null-hypothesis, which will be rejected if and only if evidence contradicts the assumptions or predictions of equilibrium ocean dynamics. Energetic constraints for equilibrium ocean dynamics allow us to put bounds on the expected circulation regimes and flow speeds, and to better understand seemingly diverging results from previous numerical simulations.

\section{Energetics of the seawater Boussinesq Equations.}

The weak compressibility of water allows us to employ the seawater Boussinesq approximation, with which the dynamical equations reduce to  \citep[e.g.][]{Young2010}
\begin{eqnarray}
\frac{D \uvec}{Dt} +2\mathbf{\Omega}\times \uvec +\nabla p &=& b\hat{k} + \mathbf{\mathcal{F}} ,  \label{momentum} \\
\nabla\cdot \uvec&=&0 ,
\end{eqnarray}
where $\mathbf{v}$ is the velocity,  $\mathbf{\Omega}$ is the planetary rotation, $p$ is the pressure anomaly relative to a hydrostatic reference state with constant density, $\rho_0$, $b=g(\rho_0-\rho)/\rho$ is buoyancy\footnote{Notice that \cite{Young2010} defines $b$ using a constant reference gravity $g_0$ and writes the first term on the R.H.S. of Eq. (\ref{momentum}) as b$\nabla Z$, where Z is $\phi/g_0$, with $\phi$ the geopotential ($g\equiv|\nabla\phi|$), such that $\nabla Z=g/g_0 \hat{k}$. We here absorb the factor $g/g_0$ into the definition of buoyancy, assuming that g is itself only a function of z.},  with $g$ the gravitational acceleration, $\hat{k}=g^{-1}\nabla{\phi}$ is the normal vector in the direction of gravitational acceleration, with $\phi$ the geopotential, and $\mathcal{F}=\mathcal{T}-\mathcal{D}$ 
is an acceleration due to tidal and frictional forces, respectively (where we define the frictional force with a negative sign for illustrative purposes, as it dominantly acts to decelerate the flow). 

Multiplying Eq. (\ref{momentum}) by $\uvec$, and using $\nabla\cdot \uvec=0$ yields an equation for the kinetic energy as
\eq
\frac{1}{2}\frac{D |\uvec|^2}{Dt} +\nabla \cdot ( \uvec p ) = wb + \uvec \cdot \mathbf{\mathcal{T}} - \uvec \cdot \mathbf{\mathcal{D}} ,
\label{KE}
\qe
where $w$ is the velocity normal to the geopotential surfaces (in practice usually well approximated by the radial velocity). 

Integrating globally over the entire ocean volume and assuming an equilibrium state and no-normal-flow boundary conditions\footnote{Notice that the no-normal flow boundary condition here amounts to neglecting any possible kinetic energy injected by jets emanating from the sea floor or ice shell \citep[e.g.][]{KiteRubin2016}. The effect of geothermal vents or freezing and melting at the ice-ocean interface on buoyant plumes, however, is included via the heat and salt flux boundary conditions.}, we find a balance between the conversion from potential to kinetic energy, $wb$, KE generation by tidal forcing, $\uvec \cdot \mathbf{\mathcal{T}}$,  and frictional dissipation, $\uvec \cdot \mathbf{\mathcal{D}}$:
\eq
\int wb dV  + \int \uvec \cdot \mathbf{\mathcal{T}} dV = \int \uvec \cdot \mathbf{\mathcal{D}} dV .
\label{global_KE}
\qe

We hence have two potential sources of kinetic energy: 1) the vertical buoyancy flux, which converts potential to kinetic energy and is directly related to the vertical heat and salt flux, and 2) tidal forcing. This paper will focus primarily on buoyancy-driven circulations, that is we assume a balance between the first term on the LHS of Eq. (\ref{global_KE}) and the  dissipation on the RHS \citep[c.f. Eq. 3.2 in][]{PaparellaYoung2002}. The potential role of tides in modulating the buoyancy-driven circulation will also be discussed. 

Importantly, Eq. (\ref{global_KE}) highlights that buoyancy forcing can energetically drive a circulation if and only if it is distributed such that the global mean advective buoyancy flux required to balance the forcing (minus any diffusive flux) is directed upwards. Notice that there is an interesting discussion in the Earth-ocean literature about the implications of this statement for a purely ``buoyancy-driven" circulation (i.e. without any mechanical forcing) in an ocean (or experimental apparatus) where all heating and cooling is applied at the surface \citep[e.g.][]{Sandstrom1908, PaparellaYoung2002, WunschFerrari2004,HughesGriffiths2008}. Energetically speaking, the energy source needed to balance any dissipation in this so-called ``horizontal-convection" problem must come from molecular diffusion (which can flux buoyancy downward to balance an upward buoyancy advection that generates KE). We will return to the potential role of molecular diffusion below. 


\subsection{Buoyancy forcing} \label{sec:forc}

The upward buoyancy flux is related to the upward flux of heat or compositional buoyancy. In this manuscript we assume a water ocean with dissolved salts, such that compositional buoyancy effects are encapsulated by the salinity. In general we thus allow buoyancy to be some function of potential temperature, salinity, and geopotential height \citep[e.g.][]{Young2010}, i.e.
\eq
b=\tilde{b}(\Theta,S,z) ,
\qe
where $\Theta$ is potential temperature\footnote{Notice that temperature is conserved following adiabatic motion in a Boussinesq fluid and hence temperature and potential temperature are formally identical. However, $\theta$ should be interpreted as potential temperature when comparing to observed fluids.}, $S$ is salinity, and $z$ is depth relative to some reference geopotential height level.  Temperature and salinity evolve according to 
\begin{eqnarray}
\label{DTdt}
\frac{D\Theta}{Dt}&=& \kappa_T \nabla^2 \Theta ,\\ 
\label{DSdt}
\frac{DS}{Dt}&=& \kappa_S \nabla^2 S ,
\end{eqnarray}
with the boundary conditions
\begin{eqnarray}
-\kappa_T \partial_z \Theta (z_{bot}) =  \frac{1}{\rho c_p} \mathcal{Q}_{bot} , &- \kappa_T \partial_z \Theta (z_{top}) &=  \frac{1}{\rho c_p}\mathcal{Q}_{top} , \\
-\kappa_S \partial_z S (z_{bot}) =  \mathcal{S}_{bot} ,    &-\kappa_S \partial_z S (z_{top}) &=  \mathcal{S}_{top} ,
\end{eqnarray}
where $\kappa_T$ and $\kappa_S$ are the molecular diffusivities of heat and salt, respectively, and $\mathcal{Q}_{bot/top}$  and $\mathcal{S}_{bot/top}$ denote the heat and salt fluxes through the bottom and top boundaries, respectively. 
For simplicity we here do not include significant heat sources in the interior, although such sources could be added. 

If the equation of state is nonlinear, the conservation equations for $\Theta$ and $S$ cannot readily be translated into a conservation equation for buoyancy. The most general way to account for nonlinearities in the equation of state, which is discussed in Appendix \ref{AppA}, is to introduce the dynamic enthalpy, which is related to the potential energy. A significantly simpler (and more intuitive) result can be obtained if we assume that horizontal variations in the thermal and haline expansion coefficients are small, such that we can approximate
\eq
b\approx \ab{b} +\ab{\alpha}(\Theta-\ab{\Theta}) + \ab{\beta}(S-\ab{S}) ,
\label{eq:b}
\qe
where $\alpha\equiv g^{-1} \partial_\theta b$ and  $\beta\equiv g^{-1} \partial_S b$ are the thermal and haline expansion coefficients and $\ab{\cdot}$ denotes a horizontal average at any given depth. Notice that $\alpha$ is typically positive (except for relatively fresh water at cold temperatures and modest pressures), while $\beta$, as defined here, is negative. 
We can then directly relate the total globally integrated upward buoyancy flux (which represents the source of kinetic energy in a buoyancy-driven flow) to the upward heat and salt fluxes: 
\eq
\B \equiv  \int wb dV \approx  \int  \left[ g \ab{\alpha}  \iint w\Theta  dA \right] dz +  \int \left[ g \ab{\beta}  \iint  wS dA \right] dz \equiv  \B^\Theta + \B^S ,
\label{B1}
\qe
where we used that $\iint w  dA=0$ due to volume conservation, and we defined $\B^\Theta $ and  $\B^S$ as the globally integrated net upward buoyancy fluxes associated with heat and salt fluxes, respectively.

\begin{figure}
\noindent\centering\includegraphics[width=0.8\textwidth]{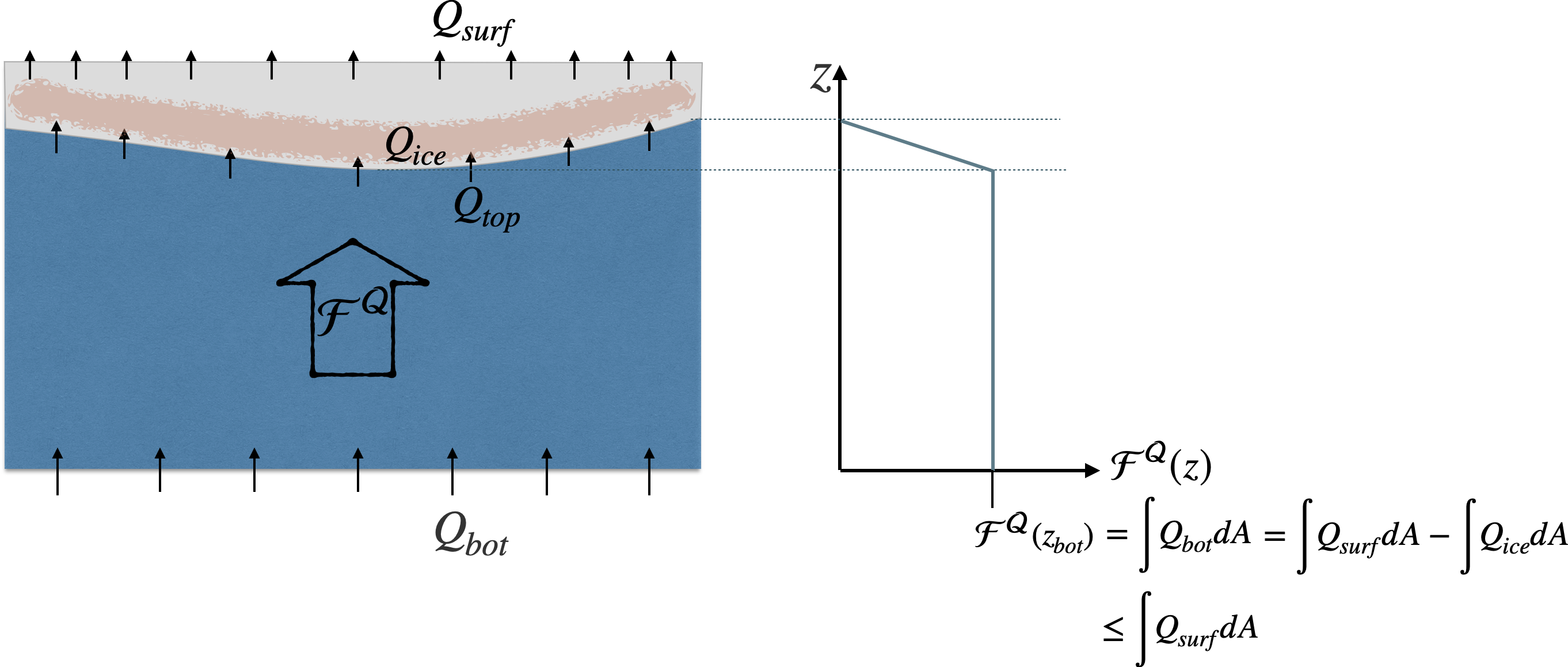}
\caption{Sketch of the vertical heat flux through the ocean, which provides the energy source for thermally driven flows. In this manuscript we assume that the ocean and ice sheet are in equilibrium, and that heat sources in the interior of the ocean are negligible, such that the globally integrated heat flux from the core ($\int Q_{bot} dA$) is equal to the net globally integrated heat flux from the ocean to the ice sheet ($\int Q_{top} dA$). Notice that the globally integrated vertical heat flux in the ocean ($\mathcal{F^Q}$) can also be constrained from the potentially more observable globally integrated heat flux emanating from the planetary body's surface ($\int Q_{surf}dA$), which is given by the sum of the core heating ($\int Q_{bot} dA$)  and the tidal energy dissipation in the ice layer ($\int Q_{ice} dA$). Specifically, the maximum globally integrated vertical heat flux in the ocean is limited to  $\mathcal{F^{Q}} \leq \int Q_{surf} dA$ .}
\label{FQ_fig}
\end{figure}

\begin{figure}
\noindent\centering\includegraphics[width=0.7\textwidth]{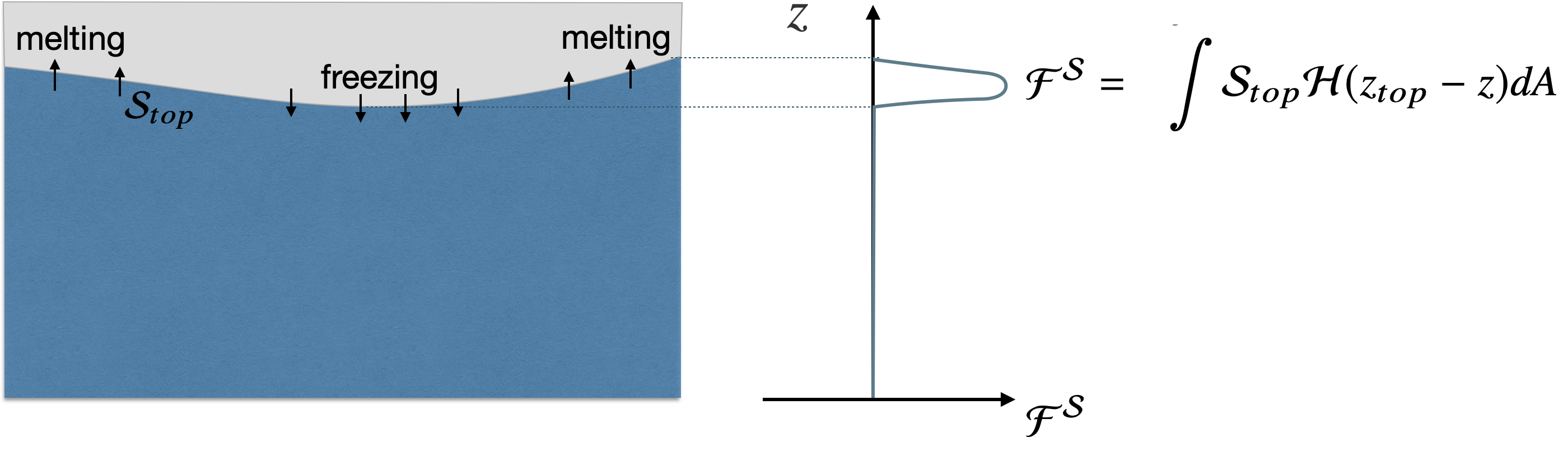}
\caption{Sketch of the vertical salt flux through the ocean. If melting occurs under thinner ice and freezing occurs under thicker ice, as sketched here, the net globally integrated salt flux required in the ocean to balance brine rejection from freezing and melting ($\mathcal{F^S}$) is upward, which implies a downward buoyancy flux that converts kinetic to potential energy. In this scenario salt fluxes therefore do not energetically drive a circulation. Instead, kinetic energy from an alternative source is required to maintain a circulation that can flux salt upwards.}
\label{FS_fig}
\end{figure}

The total upward heat and salt fluxes can be related to the sources and sinks of heat and salt via their conservation equations (see also sketches in Figs. \ref{FQ_fig} and \ref{FS_fig} ). Specifically, the area-integrated vertical advective fluxes of heat and salt are constrained by the conservation of heat and salt above or below a respective level. In a statistically steady state, where the mean rate of change of $\Theta$ and $S$ at any given depth is approximately zero, Eq. (\ref{DTdt}) can be integrated over the volume above or below some depth $z$ to obtain an equation for the vertical buoyancy flux at this depth:
\begin{eqnarray} \nonumber
\int (w\Theta  - \kappa_T \partial_z \Theta ) dA
&=& \frac{1}{\rho c_p}\left[ \int \mathcal{Q}_{bot} \heavi(z-z_{bot}) dA - \int \mathcal{Q}_{top}  \heavi(z-z_{top})\ dA \right]\\ \nonumber
&=&\frac{1}{\rho c_p}\left[ \int \mathcal{Q}_{top} \heavi(z_{top}-z) dA - \int \mathcal{Q}_{bot}  \heavi(z_{bot}-z)\ dA  \right] \\
&\equiv&\frac{1}{\rho c_p} \mathcal{F^Q} , 
\label{FQ} 
\end{eqnarray}
where $\heavi$ is the Heaviside function ($\heavi(x) = 0\;\mbox{for}\, x<0$ and $\heavi(x) = 1\;\mbox{for}\, x>0$ )  and we defined  $\mathcal{F^Q}$  as the the vertical heat flux across any depth $z$ needed to balance the net heat and salt fluxes through the boundaries above and below the respective level.
Similarly, Eq. ( (\ref{DSdt})) can be integrated to obtain
\begin{eqnarray} \nonumber
\int (wS  - \kappa_S \partial_z S ) dA&=& \int \mathcal{S}_{bot} \heavi(z-z_{bot}) dA - \int \mathcal{S}_{top}  \heavi(z-z_{top})\ dA \\ \nonumber
&=&\int \mathcal{S}_{top} \heavi(z_{top}-z) dA - \int \mathcal{S}_{bot}  \heavi(z_{bot}-z)\ dA  \\
&\equiv& \mathcal{F^S} .
\label{FS} 
\end{eqnarray}
Notice that interior sources of heat or salt could be included by modifying the definition of $\mathcal{F^Q}$ and $\mathcal{F^S}$, which generally need to balance any net sources or sinks above or below the respective level.

These definitions allow us to express the globally integrated KE generation associated with heat and salt flux  forcing as:
\eq
\B^\Theta \approx \underbrace{\int g \ab{\alpha} \frac{\mathcal{F^Q}}{\rho c_p} dz}_{\mbox{$\B^\Theta_{\mathcal{F}}$}} +  \underbrace{\int g \alpha \kappa_{T}\partial_z \theta dV}_{\mbox{$\B^\Theta_\text{diff}$}}, \quad
\label{B2_T}
\qe
and
\eq
\B^S = \underbrace{ \int g \ab{\beta} \mathcal{F^S} dz }_{\mbox{$\B^S_{\mathcal{F}}$}}  + \underbrace{ \int g \beta \kappa_{S}\partial_z S dV}_{\mbox{$\B^S_\text{diff}$}} .
\label{B2_S}
\qe
The first term on the RHS of equations (\ref{B2_T}) and (\ref{B2_S}) can be interpreted as the potential energy source associated with heat and salt forcing at the lower and upper boundaries, while the second term is the potential energy source or sink due to diffusion (which provides a source of mechanical energy whenever the stratification is statically stable, and a sink when it is unstable).

If vertical variations in $g \alpha$ and  $g \beta$ are also small, thermal forcing at the boundaries provides a source of energy that can drive a circulation if, and only if,
\eq
\alpha \int \mathcal{F}^Q dz >0  ,
\qe
i.e. for $\alpha>0$, heating, on average, needs to occur at a greater depth than cooling, while for $\alpha<0$ cooling needs to occur at greater depth than heating. Similarly salt fluxes can drive a circulation only if 
\eq
\beta \int \mathcal{F}^S dV >0 .
\qe
Since generally $\beta<0$, this requires that salt needs to be removed at greater depth and added at shallower depth.  Large vertical variations in $g \alpha$ and $g \beta$ can be accounted for using the generalized relation in Eqs. (\ref{B2_T}) and(\ref{B2_S}), which allow us to compute the mechanical energy input associated with any given heat and salt flux boundary conditions more generally. When horizontal variations in the thermal and haline expansion coefficient are non-negligible, the amount of energy that can be converted to kinetic energy depends on the specifics of the circulation, as elaborated in Appendix A.  

Eqs. (\ref{B2_T}) and (\ref{B2_S}) also highlight the potential role of vertical diffusion. Indeed, buoyancy gain and loss at the same level combined with vertical diffusion can drive a circulation, which is sometimes referred to as horizontal convection \citep[e.g.][]{Sandstrom1908, PaparellaYoung2002, WunschFerrari2004,HughesGriffiths2008}. Since the energy source in such a circulation is derived from diffusion, the energy dissipation has to go to zero in the limit of vanishing diffusivity. As a result, \cite{PaparellaYoung2002} argue that no ``turbulent" circulation can be maintained in the limit of vanishing diffusivity (and viscosity), where "turbulence"  is defined to imply a forward energy cascade with a dissipation rate that becomes independent of the viscosity\footnote{Notice that, unlike 3D turbulence, 2D (and geostrophic) ``turbulence" does generally not exhibit a forward energy cascade, making it not truly ``turbulent" by this definition. We will address constraints for geostrohic turbulence in section \ref{sec:rotation}.}. In reality, molecular diffusion is not vanishingly small and we will discuss its potential role below. It is also worth noting at this point that many numerical simulations employ strongly enhanced ``eddy diffusivities", which can act as a substantial energy source in numerical simulations. However, it is important to remember that these parameterizations are meant to represent the effect of turbulent advection---that is they parameterize the unresolved contribution to the $wb$ term in Eq. (\ref{KE}). The apparent energy source associated with parameterized ``eddy diffusion" therefore needs to be interpreted as a conversion of unresolved turbulent kinetic energy to potential energy, and accordingly can be justified only if there exists a corresponding source of unresolved turbulent kinetic energy.

\subsection{Energy dissipation and flow properties} \label{sec:diss}

In equilibrium the total sources and sinks of mechanical energy have to be in balance. The energy sources thereby provide a constraint on the energy dissipation, which in turn provides some constraint on the flow. Relating the energy dissipation rate to the kinetic energy of the flow itself is not straightforward, as it depends on the characteristics of the flow field, but we can make some progress by considering specific flow regimes and estimating the range of parameters and scales over which the flow regimes are expected to hold. In the following, we first discuss the relationship between the kinetic energy and the dissipation rate for a turbulent flow that is largely unaffected by rotation, as well as the conditions under which the assumption that rotation is negligible breaks down. We then derive an alternative relationship between the kinetic energy and the dissipation rate in the opposite limit where rotation is a leading-order effect and the most energetic flows are  geostrophically balanced (i.e. the leading order momentum balance is between the pressure gradient and Coriolis acceleration).

\subsubsection{Isotropic turbulence and the role of rotation}  \label{sec:rotation}

We first assume a fully turbulent flow unconstrained by the influence of rotation in the interior (i.e. away from the boundaries) and we assume that, in this limit, the dissipation rate is independent of the value of molecular viscosity\footnote{Notice that it is unclear in how far this holds true in the vicinity of boundaries. We here assume that most dissipation occurs in the interior in this fully turbulent regime, but we will return to the issue of boundary-layer dissipation below.}. Specifically, we start by considering Kolmogorov's theory for the kinetic energy spectrum in the inertial cascade of isotropic turbulence, which suggests that
\eq
E(k)= \mathcal{K} \epsilon^{2/3}k^{-5/3} ,
\label{Ek_Kolmogorov}
\qe
where $E(k)$ is the kinetic energy spectrum as a function of wavenumber, $k$, $\epsilon$ is the turbulent spectral KE  flux, which in turn is equal to the dissipation rate, and $\mathcal{K} \approx 1.5$ is the Kolmogorov constant \citep[e.g.][]{Vallis2006}. Integrating over the energy inertial range yields
\eq
E_t\approx  2 k_0^{-2/3} \epsilon^{2/3} ,
\label{E_Kolmogorov}
\qe
where $k_0$ is the ``injection scale" (where the KE spectrum flattens out), and $E_t$ is the total turbulent kinetic energy in the inertial range.  

We can use Eq. (\ref{E_Kolmogorov}) to estimate the kinetic energy of turbulent flows up to the largest scales of an isotropic turbulent energy inertial range. Isotropy may be broken by the geometry (e.g.  the vicinity to a boundary), or by the effects of stratification or rotation. We will here consider in particular the importance of rotation, which can break isotropy and fundamentally change the nature of the turbulent flow throughout the watercolumn.  

The effect of rotation on the turbulent flow in the interior can be characterized by the flow Rossby number, which we here define as the ratio of the inverse ``eddy turnover time-scale", $\tau_{eddy}^{-1}\sim \sqrt{E_t}k_0$, to the rotation rate, $\Omega$. This ratio measures the relative magnitude of the nonlinear advection term to the Coriolis term in the momentum equation, and thus provides a useful and widely used measure for the role of rotation in the dynamics of high-Reynolds-number flows \citep[e.g.][]{Vallis2006}. Using Eq. (\ref{E_Kolmogorov}), and $L\equiv 2\pi/k_0$, the turbulent flow Rossby number can be estimated as
\eq
Ro_t\equiv\frac{\sqrt{E_t}}{\Omega L}\approx \frac{\epsilon^{1/3}}{\Omega L^{2/3}}  .
\label{Ro_turb}
\qe
Equation (\ref{Ro_turb}) suggests a maximum length scale for turbulent flows unaffected by rotation \citep[c.f.][]{Fernando1991,JonesMarshall1993,MaxworthyNarimousa1994,Bire2022}:
\eq
L_{rot}\approx\frac{\epsilon^{1/2}}{\Omega^{3/2}}  .
\label{L_rot}
\qe
Once rotation becomes important, \cite{MaxworthyNarimousa1994} and others find that the convective Rossby number of a rotating plume in the interior scales as 
\eq
Ro_{rc}=\frac{\mathcal{\epsilon}^{1/2}}{\Omega^{3/2} H}  ,
\label{Ro_rotconv}
\qe 
where $H$ is the depth of the convecting fluid.  Eq. (\ref{Ro_rotconv}) is likely to be a better predictor of the convective flow Rossby number when ocean-depth convection is strongly affected by rotation (i.e. $L_{rot}<H$). 
If we set $L=H$, we find $Ro_{rc}= Ro_t^{3/2}$, and hence both definitions give the same prediction for the critical length/depth scale at which $Ro\approx1$ and rotation becomes important\footnote{\cite{Aubert2001} defines a parameter $ \gamma=\alpha g \mathcal{Q}_{bot} / (\rho c_p \Omega^3 H^2) = \epsilon / (\Omega^3 H^2)$ to characterize rotating convection at high Reynolds number, where in the second equality we assumed a balance between potential energy generation by the heat flux forcing and kinetic energy dissipation. With this parameter, we can write $ Ro_{t}\sim \gamma^{1/3}$ while $ Ro_{rc}\sim \gamma^{1/2}$. \cite{Aubert2001} and others (e.g. \cite{CardinOlson1994, Gastine2016}) suggest that $Ro\sim \gamma^{2/5}$  in the highly nonlinear limit of rapidly rotating convection, thus giving an intermediate power dependence of $Ro$ on $\gamma$. However, all scaling relations give the same result for the threshold where $Ro \sim 1$.}, which indeed also follows directly from dimensional analysis if we postulate that this scale shall depend only on $\epsilon$ and $\Omega$. If this length scale is significantly smaller than the scale of the most energetic flows, we expect these flows to be strongly affected by rotation. For convective flows unconstrained by rotation the natural length scale for the largest convective motion is the depth of the ocean, such that we expect the interior convective dynamics to become strongly affected by rotation if $L_{rot}$ is much smaller than the depth of the ocean \citep[c.f.][]{Fernando1991}. 


\subsubsection{Boundary layer dissipation in geostrophic dynamics}

When rotation becomes of dominant importance it is likely that much of the energy becomes trapped in geostrophically balanced vorticies and large-scale mean flows, which result from the up-scale kinetic energy cascade associated with quasi-balanced turbulent motions \citep[e.g.][]{Vallis2006}. The lack of a forward energy cascade means that dissipation is likely to be limited mostly to turbulent boundary layers near the seafloor and the ice-ocean interface. 

We can estimate the energy dissipation per unit area in a turbulent boundary layer  as \citep[e.g][and references therein]{Jansen2016}:
\eq
\int \epsilon_\text{BL} dz = c_D |U_g|^3 ,
\label{epsilon_BBL}
\qe
where the integral on the L.H.S. is over the depth of the turbulent boundary layer, $c_D$ is the turbulent drag coefficient, and $U_g$ is the characteristic near-surface geostrophic velocity. 
The value of $c_D$ depends on the surface roughness, with $c_D\approx0.0025$ an empirical average value that is commonly used for the drag coefficient for Earth's seafloor \citep[e.g.][]{Egbert2004,Sen2008}. The skin drag coefficient under smooth ice can be estimated to be around $c_D\approx0.002$, although rough morphology can significantly increase this value \citep[e.g.][]{Brenner2021}. Lacking information about the ice roughness, and noting the order-of-magnitude nature of our estimates, we will here use $c_D\approx0.001-0.01$ at both the sea floor and under the ice sheet at the top. Experience from oceanic and atmospheric modelling has shown that drag coefficients of this order produce reasonable results  for a wide range of flows \citep[e.g.][]{Smagorinsky1965,Egbert2004,Sen2008,Chen2018,Adcroft2019}, which instills confidence that a similar coefficient can be used to obtain useful order-of-magnitude estimates for icy moon oceans. 

Averaging the energy dissipation in the top and bottom boundary layers over the depth of the whole water column gives an average energy dissipation rate per unit volume 
\eq
\epsilon_\text{BL} \approx \frac{2^{5/2} c_D}{H} E_g^{3/2} ,
\label{epsilon_BBL_1}
\qe
where $H$ is the depth of the ocean and $E_g=1/2 |U_g|^2$ is the characteristic KE of the geostrophic flow. We can solve Eq. (\ref{epsilon_BBL_1}) to get a crude estimate for the characteristic  KE in a flow that is dominated by large-scale balanced dynamics, where KE dissipation occurs primarily in turbulent boundary layers:
\eq
E_g \approx \frac{H^{2/3}}{2^{5/3} c_D^{2/3}} {\epsilon_\text{BL}}^{2/3} .
\label{E_BBL}
\qe

In general, balanced dynamics may co-exist with intermediate and high Rossby number convective turbulence and/or tidal waves, in which case interior energy dissipation may be important and thus $\epsilon_\text{BL}=\epsilon-\epsilon_{int}<\epsilon$, where $\epsilon_{int}$ denotes energy dissipation away from the boundary layers. 
It is also important to note that numerical simulations of planetary circulation typically use very high artificial viscosities for numerical stability, which can lead to a large, albeit probably unphysical, dissipation of balanced KE in the interior \citep[e.g.][]{Jansen2016}. 

\section{Scaling laws for the kinetic energy of thermally-driven flows on Snowball Earth, Enceladus and Europa}\label{scalings}

We can derive estimates for the kinetic energy of buoyancy-driven flows by assuming a statistically steady state where the source of KE equals the sink. If the primary source of energy is given by thermal forcing, Eq. (\ref{global_KE})  together with Eq. (\ref{B1}) and (\ref{B2_T}) provide a scaling relation for the mean kinetic energy dissipation per unit volume:
\begin{equation}
\boxed{
\epsilon \equiv \frac{1}{V}\int \uvec \cdot\mathcal{D}  \sim \frac{\B^\theta }{V} \approx   \frac{g \alpha \mathcal{F}^Q H }{\rho c_p V}  \approx \frac{Q \alpha g}{\rho c_p} \,,
}
\label{E_source_b}
\end{equation}  
where  $Q$ is the average heat flux through the sea floor per unit area, and we for now neglect variations in the thermal expansion coefficient, $\alpha$, as well as radial variations in the gravity and the surface area throughout the depth of the ocean. (The latter assumption leads to O(10\%) errors for Europa and Enceladus, which is not of concern here. The former may lead to larger errors if the ocean is relatively fresh, and we will return to this issue later.)  We also ignored the energetic effect of molecular diffusion (the second term in Eq. (\ref{B2_T})), which is likely to be small if the ocean is convective\footnote{Notice that this formally amounts to assuming a large Nusselt number - i.e. we assume that the advective heat flux is large compared to the diffusive heat flux.}. We will return to the potential importance of molecular diffusion in a stratified ocean below.

\begin{figure}
\noindent\centering\includegraphics[width=0.5\textwidth]{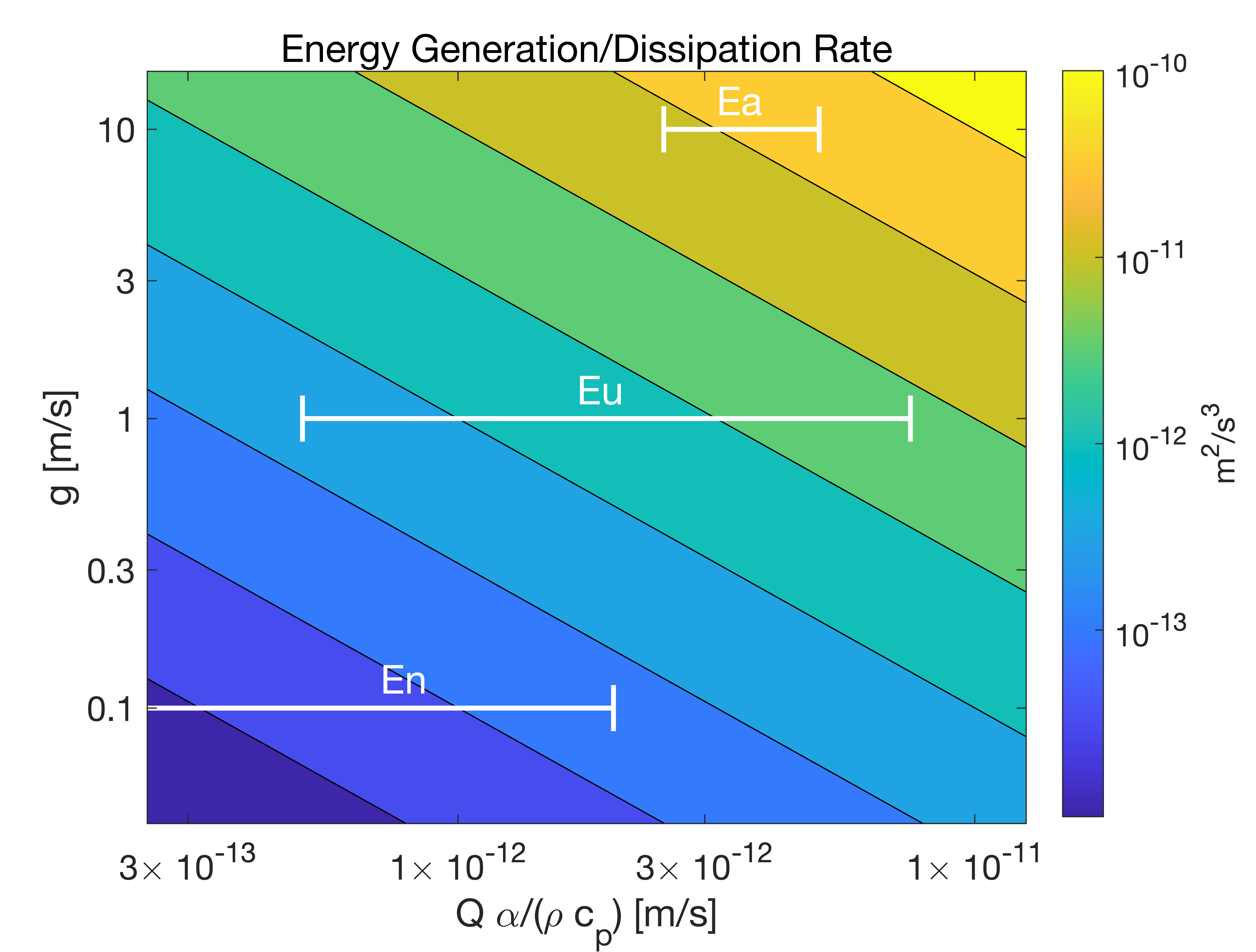}
\caption{Energy dissipation rate per unit volume for thermally-driven flows predicted by the scaling in Eq. (\ref{E_source_b}) as a function of the bottom heat flux ($Q$) times $\alpha/(\rho c_p)$ and the gravitational acceleration ($g$).  The white stars mark the estimates for Snowball Earth (Ea), Europa (Eu) and Enceladus (En), assuming the parameters given in table \ref{tab:para}. The colorbar is logarithmic with contours at $10^{-13.5\,}$m$^2$s$^{-3}$, $10^{-13\,}$m$^2$s$^{-3}$, ... , $10^{-10.5\,}$m$^2$s$^{-3}$. }
\label{epsilon_fig}
\end{figure}

Eq. (\ref{E_source_b}) provides a key constraint for the energetics of thermally-driven flows. It predicts that the energy dissipation rate per unit volume increases linearly with the bottom buoyancy flux, which in turn is given by the heat flux multiplied by the thermal expansivity and the gravity. The results are shown in Fig. \ref{epsilon_fig} for the parameter range occupied by the icy moons and Snowball Earth.  In the following, we use this result together with the results from sections \ref{sec:forc} and \ref{sec:diss} to obtain order-of-magnitude estimates for thermally-driven ocean flows on Snowball Earth, Europa, and Enceladus. 

To estimate the kinetic energy dissipation rate for thermally-driven flows in the ocean's of Snowball Earth, Europa and Enceladus, we  assume an average heating rate at the ocean floor of $Q \approx 0.1\,$W/m$^2$ for Snowball Earth \citep{Korenaga2008}, $Q\approx 0.02-0.1\,$W/m$^2$ for Europa \citep{Ruiz2005,Vance2018} and $Q\approx 0.01-0.08\,$W/m$^2$ for Enceladus \citep{Cadek2016,Choblet2017,Hemingway2018,Vance2018}. The gravitational acceleration is $g\approx 10\,$m/s$^2$ for Earth, $g\approx 1\,$m/s$^2$ for Europa and $g\approx 0.1\,$m/s$^2$ for Enceladus. 
The thermal expansion coefficient depends on temperature, salinity and pressure. Salinity in the Snowball Earth ocean was likely between around 40-70$\,$g/kg \citep[e.g.][]{Ashkenazy2013} while pressures would have ranged from $\sim 100\,$bar under the ice sheet to $\sim 400\,$bar at the sea-floor. Assuming temperatures near the freezing point, this suggests a mean ocean thermal expansivity of around $\alpha\approx1-2\e{-4}\,$K$^{-1}$. Salt concentration (and composition) on Europa are poorly constrained, and may be anywhere below about 100$\,$g/kg \citep[][]{Vance2018}. Pressures may be as low as $50\,$bar below a 5$\,$km deep ice shell while increasing to $\gtrsim 1000\,$bar at the sea floor. Assuming again temperatures near the freezing point, the mean-ocean thermal expansivity is expected to be $\alpha\approx  1 - 3 \e{-4}\,$K$^{-1}$ (although $\alpha$ may be vanishingly small at the bottom of the ice shell). Salinities on Enceladus have been estimated to be around 5-30$\,$g/kg \citep[][]{Glein2018}. Due to the relatively low pressures on Enceladus ($\lesssim 50\,$bar), the thermal expansion coefficient near the freezing point would be negative if the salinity is $\lesssim 20\,$g/kg \cite[][]{ZengJansen2021,Kang2021}, and we will return to this low-salinity scenario in section \ref{sec:low_salt}. In general, we expect $\alpha < 10^{-4}\,$K$^{-1}$ for Enceladus.  
Using further that $\rho c_p\approx 4\e{6}\,$Jm$^{-3}\,$K$^{-1}$ on all bodies, we find

\begin{eqnarray}
\epsilon & \approx & 2.5-5\e{-11} \,\mbox{m$^2$s$^{-3}$} \quad \mbox{for Snowball Earth} \\ \label{eps_snowball}
\epsilon & \approx & 5\e{-13}-8\e{-12} \,\mbox{m$^2$s$^{-3}$} \quad \mbox{for Europa} \\
\epsilon & \lesssim & 2\e{-13} \,\mbox{m$^2$s$^{-3}$} \quad \mbox{for Enceladus} \label{eps_enceladus} \,.
\end{eqnarray}

The energy dissipation rate is thus likely to be largest for Snowball Earth and smallest for Enceladus, with differences mostly driven by the differences in the gravity and further amplified by differences in the estimated vertical heat flux and thermal expansivity. For comparison, the kinetic energy dissipation rate in Earth's deep ocean today is around $2\,$TW \citep{WunschFerrari2004}, which divided by the mass of the ocean amounts to about $1.5\e{-9}\,$m$^2$s$^{-3}$. This dissipation is balanced by energy input primarily from winds and tides, which provide significantly more energy to the present-day ocean circulation than geothermal heating.

 \begin{table}
 \centering\small
 \begin{tabular}{llllll}
 \hline
                            & $Q$ [Wm$^{-2}$]  & $g$ [ms$^{-2}$] & $\alpha$ [K$^{-1}$]          & $H$ [m]              &$\Omega$ [s$^{-1}$]   \\
 \hline
  Snowball Earth  & 0.1                        & 10                        &  1$\e{-4}$ -- 2$\e{-4}$   & 2$\e3$ -- 3$\e3$     &8$\e{-5}$       \\
  Europa              &  0.02-0.1                & 1                         & 1$\e{-4}$ -- 3$\e{-4}$          & 5$\e4$ -- 1.5$\e5$   &2$\e{-5}$     \\
  Enceladus         & 0.01-0.08               &0.1                       &  $<1 \e{-4}$                      & 1$\e4$ -- 5$\e4$      &5$\e{-5}$       \\
  \hline
 \end{tabular}
 \caption{Overview of assumed parameter ranges for the sea floor heating rate, $Q$, gravitational acceleration, $g$, thermal expansion coefficient, $\alpha$, ocean depth, $H$, and Coriolis parameter, $f$, for Snowball Earth, Europa, and Enceladus. (Sources are cited in the text.) Notice that uncertainties in gravity and rotation rate are comparatively small and hence ignored here.}
 \label{tab:para}
 \end{table}

 \begin{table}
 \centering\small
 \begin{tabular}{lllll}
 \hline
                             &$\epsilon$ [m$^2$s$^{-3}$]  &$L_{rot}$[m]     & $Ro_{rc}$              &$E_{g}$[m$^2$s$^{-2}$] \\
 \hline
  Snowball Earth    & $2.5-5\e{-11}$                       & $7-10$          & $2-5\e{-3}$             & $10^{-4}-10^{-3}$  \\
  Europa                & $5\e{-13}-8\e{-12}$                & $8-30$          & $1-8\e{-4}$             & $5\e{-5}-3\e{-3}$  \label{epsilon_europa} \\ 
  Enceladus           & $\lesssim 2\e{-13} $               &$\lesssim 1$  & $\lesssim 1\e{-4} $ & $\lesssim 10^{-4}$ \\
  \hline
 \end{tabular}
 \caption{Overview of estimates for mechanical energy input (per unit mass) due to thermal forcing, $\epsilon=\B/H$, the critical length scale where rotation is expected to become important, $L_{rot}$, the convective Rossby number for rotating convection, $Ro_{rc}$, and an estimate of the maximum characteristic KE energy (assuming small interior dissipation) of the large-scale geostrophic flow, $E_{g}$, for Snowball Earth, Europa, and Enceladus.}
 \label{tab:results}
 \end{table}

\subsection{The role of rotation}

\begin{figure}
\noindent\centering\includegraphics[width=0.5\textwidth]{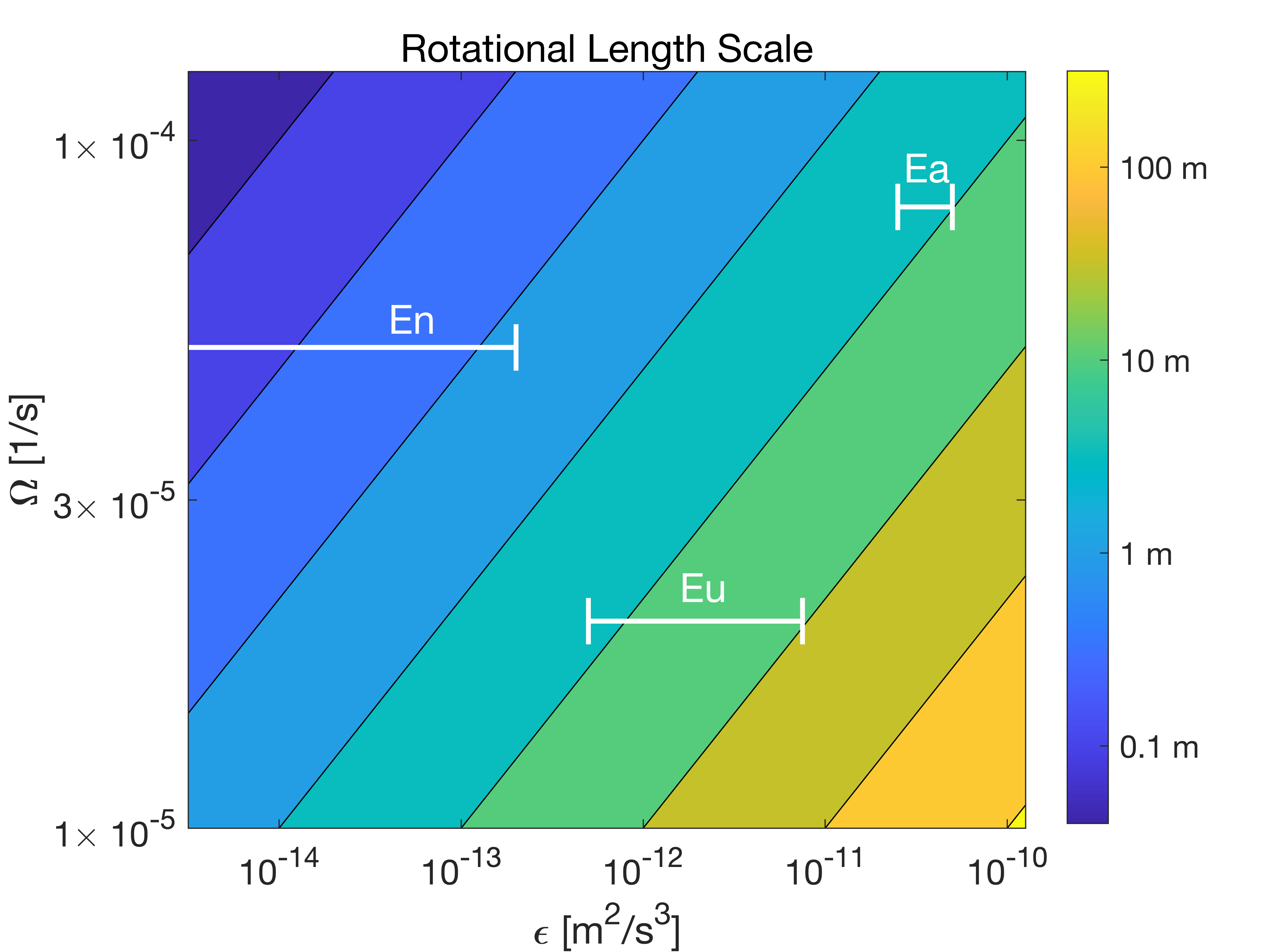}
\caption{Estimated length scale above which turbulence in the interior is expected to become strongly affected by rotation, as predicted by the scaling in Eq. (\ref{L_rot}), as a function of the turbulent dissipation rate ($\epsilon$) and the rotation rate ($\Omega$).  The white bars mark the estimates for thermally-driven circulation in Snowball Earth (Ea), Europa (Eu) and Enceladus (En), assuming the parameter ranges given in table \ref{tab:para}. The markers indicating Snowball Earth, Europa and Enceladus further assume that most of the turbulent kinetic energy dissipation occurs in the interior. If the dissipation is dominated by boundary layers, the markers would move to the left, thus further reducing the length scale where the interior flow would become strongly affected by rotation. The colorbar is logarithmic with contours at $10^{-1\,}$m, $10^{-0.5\,}$m, ..., $10^{2.5\,}$m }
\label{L_rot_fig}
\end{figure}

Given the energy dissipation rate, $\epsilon$, and the rotation rate of the planetary body, $\Omega$, we can use Eq. (\ref{L_rot}) to estimate a critical scale $L_{rot}$ above which we may expect turbulent motions to become strongly affected by rotation (Fig. \ref{L_rot_fig}). Using  the energy dissipation rates estimated above and $\Omega\approx 8\e{-5}\,$s$^{-1}$ for Snowball Earth,  $ 5\e{-5}\,$s$^{-1}$ for Enceladus, and $2\e{-5}\,$s$^{-1}$ for Europa, we find the scale at which convection is expected to become strongly affected by rotation from Eq. (\ref{L_rot}) to be $L_{rot} \sim$ O(10$\,$m) for Snowball Earth and Europa and $L_{rot} \lesssim 1\,$m for Enceladus.
In addition to parametric uncertainties (quantified in Fig. \ref{L_rot_fig} and table \ref{tab:results}), it is possible that assumptions leading to Eq. (\ref{L_rot}) may not hold. In particular, we here assumed that the interior dissipation (which enters the scaling for $L_{rot}$) is of similar order as the total kinetic energy dissipation. If the dissipation is dominated by boundary layers, the length scale where the interior flow would become strongly affected by rotation would be further reduced. Regardless, $L_{rot}$ is much smaller than the ocean depth in all three cases and we therefore expect deep convection and any other potential large-scale flow to be strongly affected by rotation. It is also worth noting that for Snowball Earth and Europa  $L_{rot}$ is likely to be much larger than the Kolmogorov scale where viscous dissipation occurs, $L_{\nu}=(\nu^3/\epsilon)^{1/4}$, which is expected to be on the order of a few cm, thus indicating that an isotropic inertial range is expected to exist at scales smaller than $L_{rot}$.  On Enceladus it is not clear whether $L_{rot}$ is significantly larger than the Kolmogorov scale, raising the possibility that no isotropic cascade exists. In that case, the relationship in Eq. (\ref{Ek_Kolmogorov}) may not be expected to hold at any scale, but the conclusion that the large-scale flow would be strongly affect by rotation would remain true, as indeed the flow at all scales (down to the viscous dissipation scale) would be strongly affected by rotation. We also note that, for all icy oceans considered here, $L_{rot}$  is significantly smaller than the smallest length scales that can be resolved in global-scale numerical simulations of these oceans, suggesting that ``large-eddy-simulations" that can explicitly resolve part of the isotropic turbulent inertial range are not feasible with realistic parameters (c.f. \cite{Bire2022}).  

\subsubsection*{Comparison to existing simulation results}

\begin{figure}
\noindent\centering\hspace{-0.085\textwidth}\includegraphics[width=1.15\textwidth]{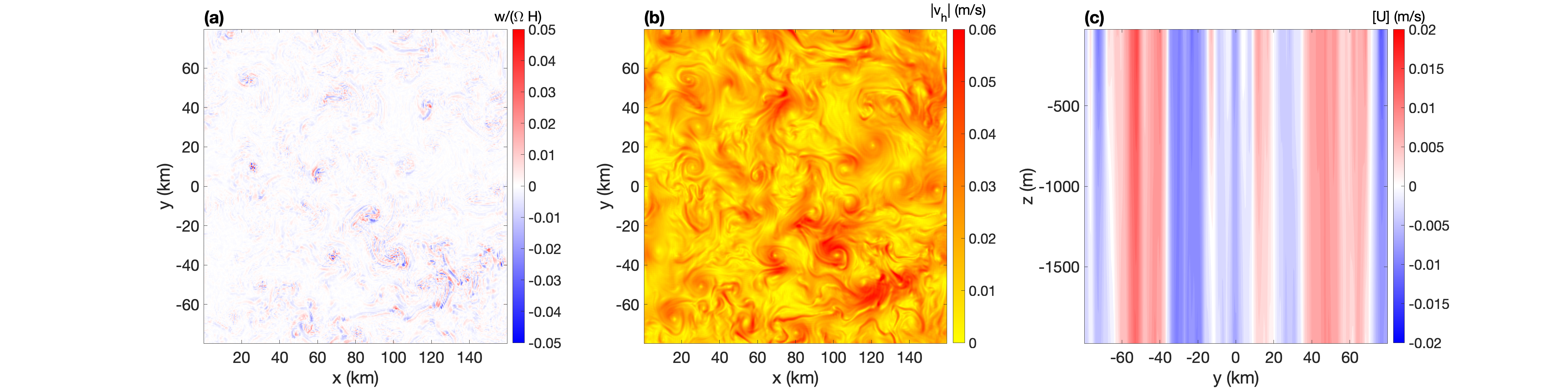}
\caption{Results from an idealized Snowball Earth ocean simulation, originally presented in \cite{Jansen2016}. \textbf{a)} Snapshot of the non-dimensional vertical velocity normalized as $w/(\omega H)$, where $\omega$ is the planetary rotation rate and $H$ is ocean depth, such as to provide a flow Rossby number. \textbf{b)} Snapshot of the horizontal flow speed near the sea floor. \textbf{c)} Snapshot of the zonal mean zonal flow (as a function of depth and latitude). The model simulates buoyancy-driven flow associated with a spatially inhomogeneous bottom heating with a mean of 0.1 W/m$^2$, crudely mimicking the expected dynamics in a Snowball-Earth Ocean. The simulation uses a linear equation of state with thermal expansivity $\alpha=1\e{-4}$K$^{-1}$ and an idealized cartesian-coordinate domain representative of the mid latitudes \citep[see][for additional details]{Jansen2016}. }
\label{Snowball}
\end{figure}

\begin{figure}
\noindent\centering\hspace{-0.08\textwidth}\includegraphics[width=1.15\textwidth]{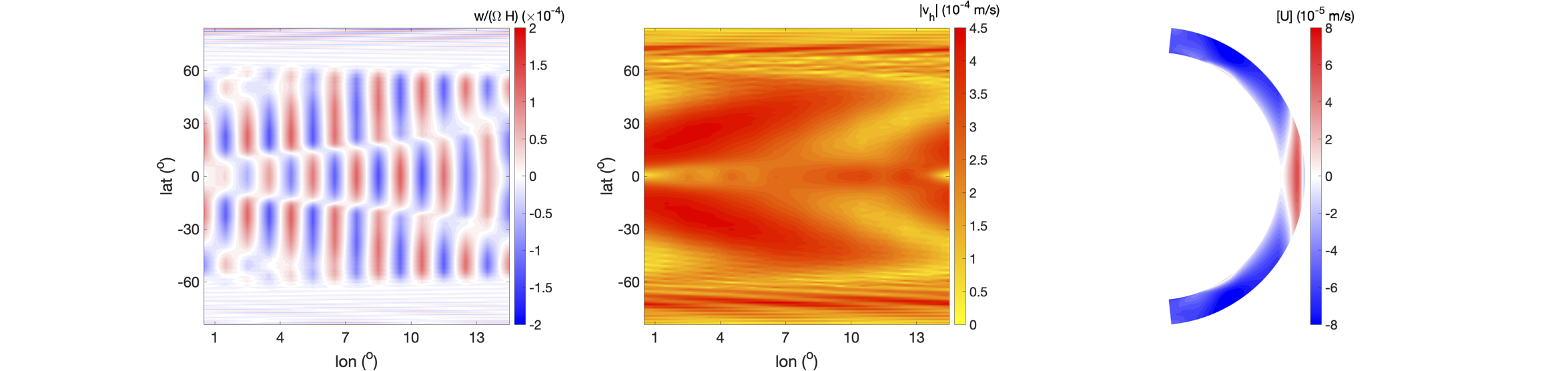}
\caption{Results from an idealized high-salinity Enceladus ocean simulation, originally presented in \cite{ZengJansen2021}. \textbf{a)} Snapshot of the non-dimensional vertical velocity normalized as $w/(\omega H)$, where $\omega$ is the planetary rotation rate and $H$ is ocean depth, such as to provide a flow Rossby number. \textbf{b)} Snapshot of the horizontal flow speed near the sea floor. \textbf{c)} Snapshot of the zonal mean zonal flow (as a function of depth and latitude). The model simulates buoyancy-driven flow associated with a spatially inhomogeneous bottom heating with a mean of 0.04 W/m$^2$, crudely mimicking the conditions in Enceladus' Ocean. The simulation assumes a spatially constant salinity of $35$g/kg with an Earth-like salt composition and simulates the flow in a $15$-degree wide sector of a spherical shell \citep[see][for additional details]{ZengJansen2021}. }
\label{Enceladus_hs}
\end{figure}

The importance of rotation in the oceans of Snowball Earth, Europa, and Enceladus is qualitatively consistent with the Snowball Earth simulations of \cite{Ashkenazy2013} and \cite{Jansen2016}, the Europa simulations of \cite{AshkenazyTziperman2020} and the Enceladus simulations of \cite{Kang2020,Kang2021} and \cite{ZengJansen2021}.  For illustration, Figures \ref{Snowball} and \ref{Enceladus_hs}, show simulation results from \cite{Jansen2016} and \cite{ZengJansen2021} for heating-driven flows in the oceans of Snowball Earth and Enceladus, respectively. 

The flow field in the Snowball-Earth simulation shows a horizontal eddy field characteristic of geostrophic turbulence (Fig. \ref{Snowball}b), with the flow largely barotropized - i.e. varying little in the vertical (Fig. \ref{Snowball}b). The characteristic vertical flow Rossby numbers are small (Fig. \ref{Snowball}a), qualitatively consistent with the predictions in section \ref{sec:rotation}. However, we notice that the vertical flows here are associated with the quasi-geostrophic eddies rather than convection, as geostrophic eddies establish a statically stable stratification. Neither of the convective Rossby-number scalings in Eqs. (\ref{Ro_turb}) and (\ref{Ro_rotconv}) are therefore expected to apply to this flow (although the conclusion that rotation is important still holds, as it is indeed a necessary condition for the development of quasi-geostrophic eddies in the first place).

The vertical flow Rossby numbers in the high-salinity Enceladus Ocean simulations of \cite{ZengJansen2021} are even smaller, and show a pattern of grid-scale convection (Fig. \ref{Enceladus_hs}a). The grid-scale convection is qualitatively consistent with the prediction that the scale at which rotation affects convective plumes (O(1m)) is much smaller than the model resolution ($\sim$1km), but unfortunately also implies that the details of the simulated flow are expected to be sensitive to the model resolution and somewhat arbitrary modeling choices, such as parameterized ``eddy" viscosities and diffusivities \citep[c.f.][]{ZengJansen2021}. Nevertheless, the importance of rotation is robust and can also be seen in the organization of the zonal mean flow into jets that are approximately aligned with the axes of rotation (Fig. \ref{Enceladus_hs}c).

The result that flows are very strongly affected by rotation may appear at odds with simulation results for Europa's ocean by \cite{Soderlund2014}, which show strong turbulent convection with relatively moderate flow Rossby numbers. The apparent contradiction can readily be understood by noting that the dimensional vertical heat flux in the DNS of \cite{Soderlund2014} would amount to about $10^5\,$Wm$^{-2}$ after re-dimensionalizing with Europa's planetary parameters---versus a heat flux of $0.02-0.1\,$Wm$^{-2}$ assumed here. Increasing the vertical heat flux on Europa to $10^5\,$Wm$^{-2}$ would yield $L_{rot}\approx 25\,$km, which in turn would make significant turbulent convection that is only relatively weakly affected by rotation plausible. Indeed, using Eq. (\ref{Ro_turb}) with $L\approx H\approx100\,$km (for the depth of Europa's ocean) and a heat flux of $10^5\,$Wm$^{-2}$ (which gives $\epsilon\approx5\e{-6}\,$m$^{2}$s$^{-3}$) we can estimate a Rossby number for full-depth turbulent convection  of $Ro_t\approx 0.4$, which is broadly consistent with the Rossby numbers of convective motions found in the simulations of \cite{Soderlund2014}. The large implied dimensional heat flux in the DNS of  \cite{Soderlund2014} is meant to account for the large viscosities and diffusivities that are necessary in numerical simulations. However, the energetic argument discussed here clarifies that the greatly amplified heat flux is expected to lead to much larger dimensional flow speeds. The magnitude of the dimensionalized velocities from the DNS of \cite{Soderlund2014}, and the associated large-scale flow Rossby numbers, hence should not be taken at face value.

In summary, our results suggest that the turbulent flow characteristics in the ocean's interior are strongly constrained by the rate of energy input, which in the case of themally-driven flows is given by the vertical heat flux, and the nature of the energy dissipation. The dissipation mechanisms and in some cases the energy source are likely to be misrepresented in numerical modeling studies, which cannot resolve all relevant scales of motion. The quantitative results of icy moon ocean  simulations therefore need to be interpreted with care.  

\subsection{Quasi-balanced flows} 

\begin{figure}
\noindent\centering\includegraphics[width=0.5\textwidth]{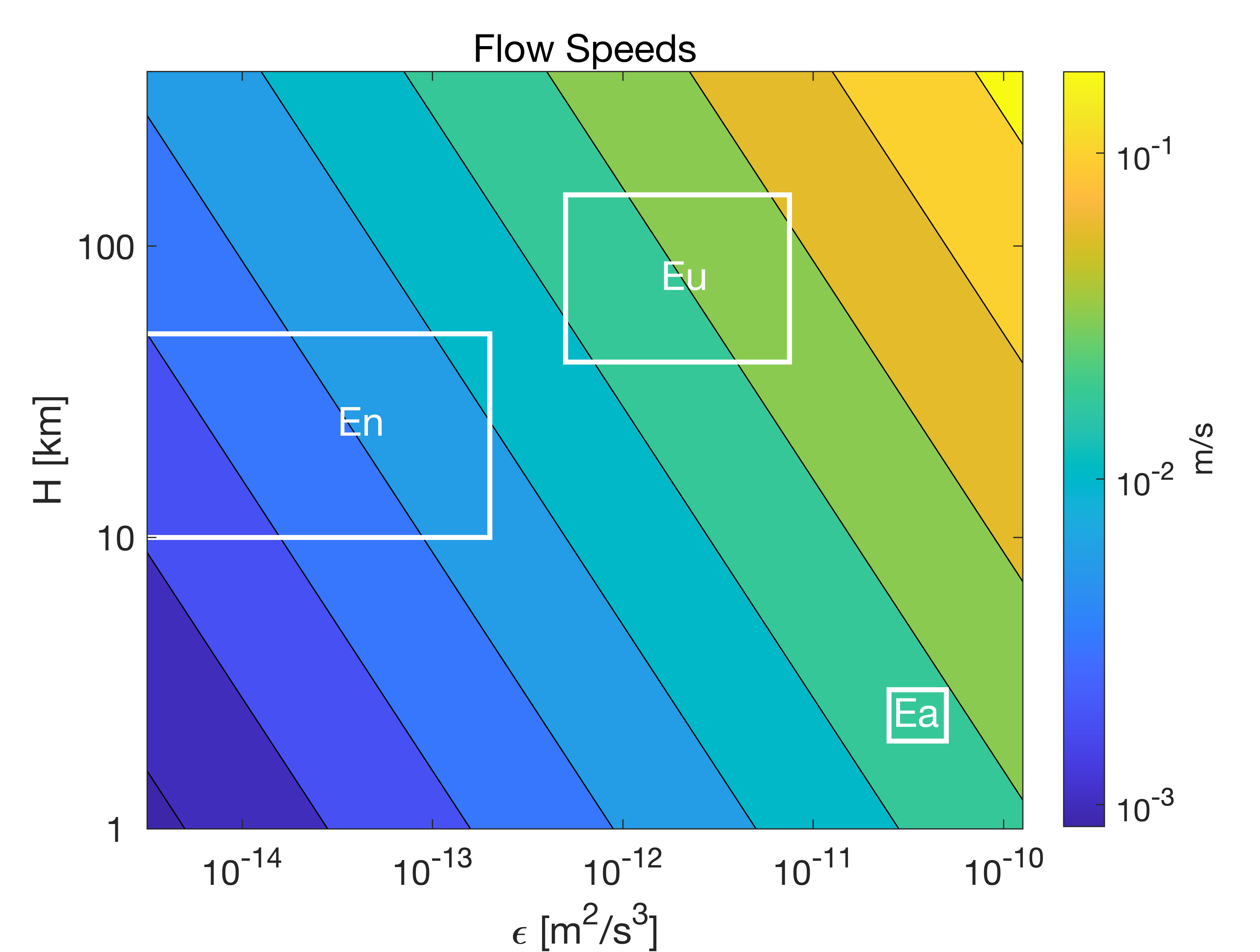}
\caption{Estimated geostrophic flow speeds ($U_g=\sqrt{2E_g}$) as a function of the energy dissipation rate ($\epsilon$) and ocean depth ($H$) as predicted by the scaling in Eq. (\ref{E_BBL}) assuming that most dissipation happens in turbulent boundary layers near the sea floor and ice-ocean interface with a drag coefficient $C_d=0.025$. Notice that a one order of magnitude uncertainty in $C_d$ would amount to an uncertainty in the flow speed of just over a factor of two ($10^{1/3}$). The white boxes mark the estimated locations for Snowball Earth (Ea), Europa (Eu) and Enceladus (En), assuming the parameter ranges given in table \ref{tab:para}. The colorbar is logarithmic with contours at $10^{-3\,}$m$\,$s$^{-1}$, $10^{-2.75\,}$m$\,$s$^{-1}$, ... , $10^{-0.75\,}$m$\,$s$^{-1}$.}
\label{KE_fig}
\end{figure}

At scales much larger than $L_{rot}$, we expect the motion to be strongly affected by rotation, leading to quasi-balanced flows that do not undergo a forward energy cascade. Energy dissipation then may be limited mostly to turbulent boundary layers, such that we can assume $\epsilon_\text{BL}\sim\epsilon$ . In this case, Eq (\ref{E_BBL}) may provide a useful estimate for the total KE and thus typical flow speeds as a function of the energy dissipation rate, $\epsilon$,  and the depth of the ocean, $H$ (Fig. \ref{KE_fig}).   Assuming an ocean depth of  $H\approx2-3\,$km for Snowball Earth \citep{Ashkenazy2013,Yang2017},  $H\approx50-150\,$km for Europa \citep[][]{Anderson1998,Vance2018} and $H\approx10-50\,$km for Enceladus \citep{Cadek2016,Hemingway2018,Vance2018}, we find 
\begin{eqnarray}
E_g & \approx & 10^{-4}-10^{-3}\,\mbox{m$^2$s$^{-2}$} \quad \mbox{for Snowball Earth} \label{Eg_Snowball} \\
E_g & \approx & 5\e{-5}-3\e{-3}\,\mbox{m$^2$s$^{-2}$} \quad \mbox{for Europa} \\
E_g & \lesssim &10^{-4}\,\mbox{m$^2$s$^{-2}$} \quad \mbox{for Enceladus.} \label{Eg_Enceladus}
\end{eqnarray}
These results suggest potential balanced flow velocities of up to a few cm/s for Snowball Earth and Europa and up to one cm/s for Enceladus. These estimates should generally be viewed as upper bounds as we here assumed that all energy input goes into balanced motions that are dissipated only in the boundary layers, while we ignored any potential additional dissipation in the interior, which would reduce the expected kinetic energy level. 

\subsubsection*{Comparison to  existing simulation results  }

\begin{figure}
\noindent\centering\includegraphics[width=\textwidth]{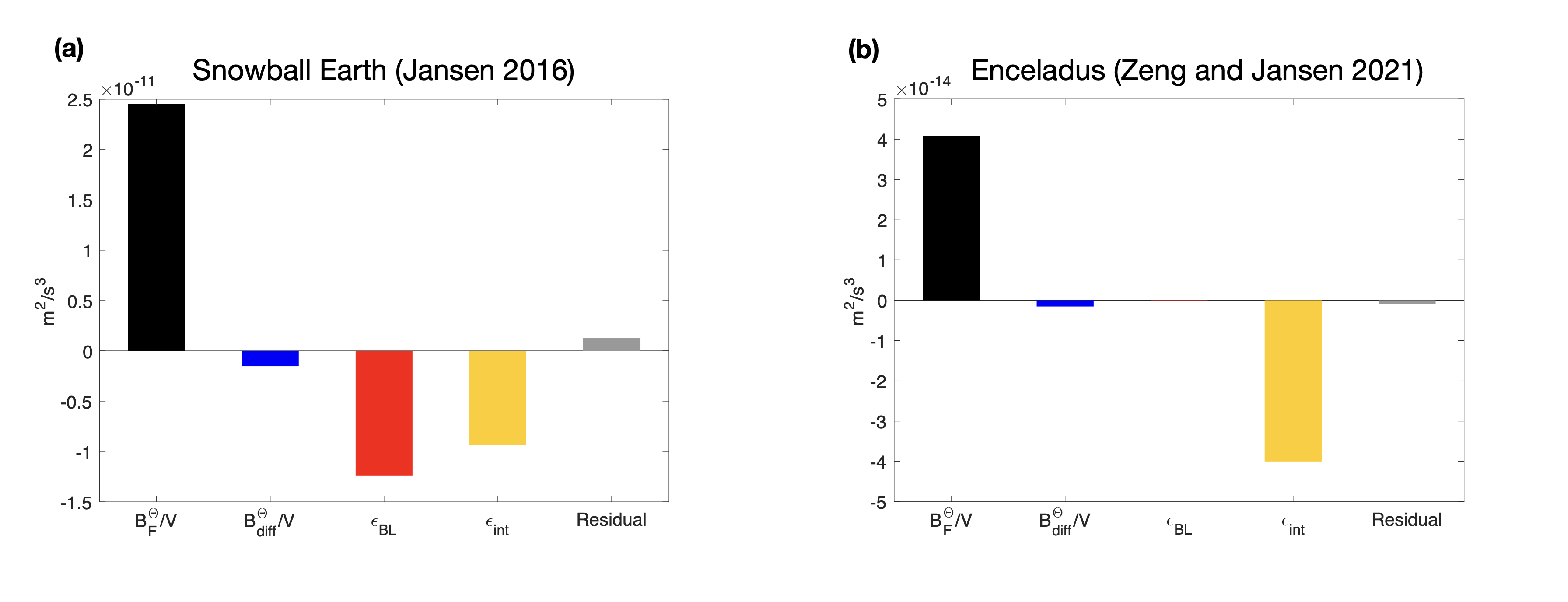}
\caption{Mechanical Energy budget of \textbf{(a)} Snowball Earth ocean simulation from \cite{Jansen2016} and \textbf{(b)} Enceladus ocean simulation from \cite{ZengJansen2021}. From left to right are the the energy sources and sinks (per unit volume) due to heat flux forcing at the boundaries ($\B^\Theta_{\mathcal{F}}/V$), vertical diffusion  ($\B^\Theta_\text{diff}/V$), parameterized boundary layer dissipation ($\epsilon_\text{BL}$) and interior viscous dissipation ($\epsilon_\text{int}$). The last bar shows the residual associated with non-equilibrium effects and errors in our offline calculation of the energy budgets.}
\label{Energetics_Snowball_Enceladus}
\end{figure}

Flow velocities on the order of centimeters per second are broadly consistent with the Snowball Earth simulations of \cite{Jansen2016} (Fig. \ref{Snowball}b), as well as the Snowball Earth simulations of \cite{Ashkenazy2013} and Europa simulations of \cite{AshkenazyTziperman2020}. Figure (\ref{Energetics_Snowball_Enceladus}a) shows the energy budget terms for the Snowball Earth simulation of \cite{Jansen2016} (c.f. Fig. \ref{Snowball}b for the corresponding flow fields). As assumed in our scaling argument, the primary energy source is associated with the vertical heat flux imposed by the sea floor heating, whose magnitude $\B^\Theta_{\mathcal{F}}/V\sim \epsilon \approx 2.5\e{-11}$m$^2$s$^{-3}$ is consistent with our estimate in Eq. (\ref{eps_snowball}). Somewhat more than half of this kinetic energy is dissipated at the boundary (where the model uses a quadratic drag, consistent with the assumption in Eq. \ref{epsilon_BBL}), with the remainder dissipated by viscous dissipation in the interior. It is likely that the interior dissipation is unrealistically large in the simulations due to the limited resolution and required large ``eddy" viscosity. However, the interior dissipation does not affect the order-of-magnitude of the velocity as long as boundary friction (assumed to be dominant in the derivation of Eq. (\ref{E_BBL})) is a first order contributor to dissipation, and hence the simulated velocities are consistent with the prediction in Eq. (\ref{Eg_Snowball}).  

The high-salinity Enceladus Ocean simulation of \cite{ZengJansen2021} has 
flow velocities on the order of $10^{-4}\,$m/s (Fig. \ref{Enceladus_hs}), which is broadly similar results have been reported in \cite{Kang2020}, but two orders of magnitude weaker than our estimate for the maximum flow speeds (Eq. \ref{Eg_Enceladus}). The parameters used in the Enceladus simulation of \citet{ZengJansen2021} are expected to place the energy input about a factor of 10 below our upper bound estimate, which by itself would translate to flow speeds that are only about a factor of 2 below the predicted maximum (c.f Eq. \ref{epsilon_BBL}).  Some discrepancy may be expected due to the parameterized boundary-layer drag.  Instead of the quadratic boundary layer drag assumed in Eq. (\ref{epsilon_BBL_1}), the simulations of \cite{ZengJansen2021} (and also \cite{Kang2020}) apply a relatively strong linear drag parameterization. 
However, the stronger boundary layer drag (which can readily be incorporated into the scaling law) can only explain a relatively small part of the misfit. Instead, interior viscous dissipation is key to explaining the much weaker velocities in the simulations, as shown explicitly in Fig. \ref{Energetics_Snowball_Enceladus}b for the Enceladus simulation of \citet{ZengJansen2021}.  
Strong dissipation of kinetic energy associated with the (numerically necessary) large interior viscosity implies that the total dissipation rate is much larger than the boundary layer dissipation( i.e. $\epsilon\gg \epsilon_{BL}$), thus leading to much weaker velocities. The large viscosities in the simulations (which cannot resolve the inertial cascade range) cannot be justified on physical grounds. Whether a dominant contribution from interior dissipation is likely to exist on Enceladus, or is purely an artifact of insufficient resolution and artificially large viscosity in the simulations, remains unknown and depends on how efficiently energy is transferred into balanced motions. In either case, the estimate in Eq. (\ref{Eg_Enceladus}) is expected to remain valid as an upper bound, although the true velocities may be much smaller. 

The Europa simulations of \cite{Soderlund2014} as well as the Enceladus-like small Ekman number simulations of \cite{Soderlund2019} suggest much larger velocities than estimated here (if re-dimensionalized with the actual ocean depth and rotation rate of the respective icy moons). Qualitatively, this discrepancy may be expected as a result of the much larger dimensional vertical heat flux, which leads to a much larger energy source, as discussed above. However, the scaling argument leading to Eq. (\ref{E_BBL}) is not expected to apply to the simulations of \cite{Soderlund2014} and \cite{Soderlund2019}, due to different boundary conditions, thus not allowing for a quantitative prediction of their simulation results. Specifically,  \cite{Soderlund2014} and \cite{Soderlund2019} use smooth boundaries with free-slip boundary conditions, which implies no boundary drag. 
 
\subsection{Thermally-driven flows in a low-salinity, low-pressure ocean} \label{sec:low_salt}

\begin{figure}
\noindent\centering\includegraphics[width=\textwidth]{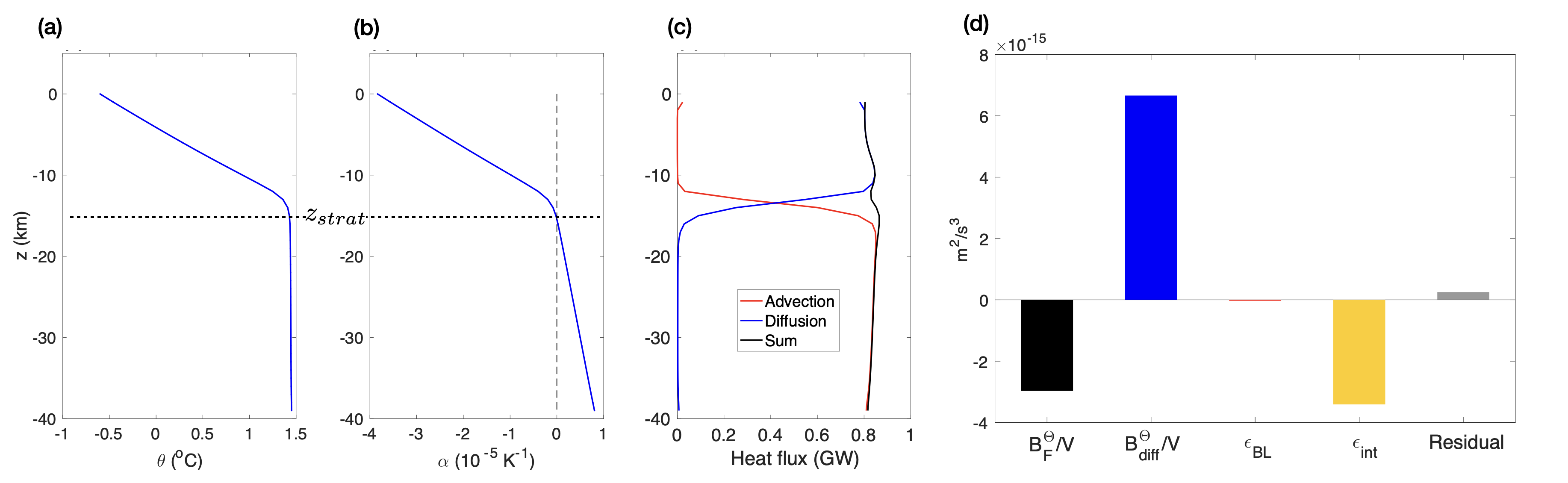}
\caption{Results from Enceladus Ocean simulation with low salinity from \citet{ZengJansen2021} (as described in Fig. 6 but with salinity 8.5 g/kg).  Although the ocean is heated from below, convection is suppressed in the upper ocean as the thermal expansivity is negative near the freezing point. \textbf{(a)}  temperature, \textbf{(b)} thermal expansivity, \textbf{(c)} advective (red), diffusive (blue) and total (black) vertical heat flux, all as a function of depth. Panel \textbf{(d)} shows the energy budget decomposition as in Fig. \ref{Energetics_Snowball_Enceladus}. Panels a-c are reproduced from \citet{ZengJansen2021}.} 
\label{Enceladus_LS}
\end{figure}

If salinity and pressure are relatively low, the thermal expansion coefficient near the freezing point is negative. This scenario was first suggested for Europa by \cite{Melosh2004}, although due to the modest pressures that are required for a negative thermal expansion coefficient, it appears most likely to be relevant for Enceladus, and was considered in the low-salinity Enceladus ocean simulations of \cite{ZengJansen2021} and \cite{Kang2021}. 
To estimate the energetics of such an ocean, we assume a stably stratified layer below the ice sheet and above some depth $z_{strat}$, as found in the low-salinity simulation of \cite{ZengJansen2021} (Fig. \ref{Enceladus_LS}). In the stratified layer, the temperature decreases upwards from the ``critical temperature", $\Theta_c$, where $\alpha(\Theta_c,z_{strat})=0$ to the freezing point, $\Theta_f$, at the bottom of the ice sheet, $z_{ice}$ (where $\alpha(\Theta_f,z_{ice})<0$). For illustrative purposes, we assume that the sea floor and ice sheet are flat (i.e. both boundaries follow a geopotential height surface), such that $\mathcal{F^Q}$ is vertically constant and equal to the total bottom heat flux. Assuming again that horizontal variations in the thermal expansion coefficient are small, Eq. (\ref{B2_T}) gives the KE generation as
\eq
\B^\Theta = \int g \ab{\alpha} \left[  \frac{\mathcal{F^Q}}{\rho c_p} + A\kappa_{T}\partial_z \ab{\theta} \right]dz ,
\label{B_T_Enc} 
\qe
In the stratified layer, $\ab{\alpha}<0$, such that the first term in the integral in (\ref{B_T_Enc}) amounts to an energy sink. One plausible solution is then that the vertical heat flux through the stratified layer is balanced by molecular diffusion (i.e. the second term in the integral in \ref{B_T_Enc}), which amounts to a conversion from internal to potential energy. In this case the stratified layer would only be a few tens of meters thick \citep{ZengJansen2021}, and most of the ocean would be convective with  $\Theta>\Theta_c$ and hence $\alpha>0$. The dynamics in the convective layer would follow scaling laws analogous to those presented above, although the very small thermal expansion coefficient at temperatures only marginally warmer than $\Theta_c$ implies that the energy input and hence maximum flow speeds would be much smaller than the upper bound estimated above.  


The simulations of \cite{ZengJansen2021} use a relatively large ``eddy diffusivity" ($\kappa_z=5\e{-5}$m$^2$s$^{-1}$), chosen such as to obtain a well resolved stratified layer with negative thermal expansivity, above a convective deep ocean with weakly positive thermal expansivity (Fig. \ref{Enceladus_LS}a/b). The vertical heat flux through the stratified layer is balanced by this ``eddy diffusion" (Fig. \ref{Enceladus_LS}c), such that from the perspective of the model's energetics, the two terms in the integral in (\ref{B_T_Enc}) balance in the stratified layer. Below the stratified layer, the thermal expansivity is positive such that convective upward heat flux releases kinetic energy that can balance dissipation. Integrated over the entire ocean, however, diffusion is the only net energy source and balances the energy loss associated with net upward heat flux and the viscous dissipation (Fig. \ref{Enceladus_LS}d).  The simulations of \cite{Kang2021} use an even larger eddy diffusivity ($\kappa=5\e{-3}$m$^2$s$^{-1}$), such that the entire depth of the ocean becomes stratified with the upward heat flux and associated downward buoyancy flux accomplished by ``eddy diffusion" (not shown). However, in both studies the model's diffusivities, which are much larger than the molecular value, need to be interpreted as representing turbulent mixing. If these turbulent mixing rates are real, the associated vertical heat flux represents turbulent advection, which amounts to a negative  $\B^\Theta$ - that is a conversion of turbulent kinetic energy to large-scale potential energy.  An additional implied energy source is therefore needed to generate the turbulent kinetic energy, and justify the applied eddy diffusivities. A possible source are tides and/or librations, to which we will return below.

\section{Salinity Forcing}

A number of studies have recently argued for the importance of salt forcing in driving a circulation on Europa and Enceladus \citep{Zhu2017, Kang2020, AshkenazyTziperman2020, Kang2021, Lobo2021}. Since $\beta$, as defined in Eq. (\ref{eq:b}), is always negative, salinity forcing provides a source of mechanical energy if, and only if, salt is added (e.g. by freezing) at a shallower depth and removed (or freshwater is added by melting) at a greater depth (Eqs. \ref{FS} and \ref{B2_S}). At steady state, however, freezing and melting need to balance the convergence of ice flow, which tends to flow from regions of thick ice to regions of thin ice. We therefore expect melting to occur where the ice is thin, and hence the ice-ocean interface is shallow, while freezing occurs where the ice is thick and hence the ice-ocean interface is deep (c.f. Fig. \ref{FS_fig}). In this case the required upward salt flux reduces the kinetic energy of the ocean and hence cannot, energetically speaking, drive a circulation.

Although salt and freshwater flux forcing is not likely to represent a source of energy, it can generate a stable stratification over much of the ocean, which in turn allows diffusion to transport fresh buoyant water downward, thus allowing for the maintenance of a diffusively-driven overturning circulation (via the second term in Eq. \ref{B2_S}). We can estimate an upper bound for the energy source associated with molecular salt diffusion, given a maximum vertical salinity contrast. In an ocean driven by surface freshwater forcing, most of the ocean is expected to fill up with the saltiest water, leaving a relatively fresh surface layer in the region of net melting. The vertical salt contrast can then be approximately bounded by the global mean ocean salinity, as the deep ocean's salinity is expected to be near this mean value, while the surface salinity may be smaller but has to be positive.  This gives an estimated upper bound for the global mean energy source rate  as
\begin{equation}
\epsilon \lesssim \frac{\kappa_S g \beta S}{H},
\label{saltdiff}
\end{equation}
For Snowball Earth,  using $\kappa_S\sim 10^{-9}\,$m$^2$/s,  $g=10\,$m/s$^2$, $\beta\sim 10^{-3}\,$kg/g, $S\lesssim 70\,$g/kg  and $H\gtrsim 2\e{3}\,$m, we get $\epsilon\lesssim 3\e{-13}\,$m$^2$/s$^3$.  For Europa, using $S\lesssim 100\,$g/kg, $g=1\,$m/s$^2$, and $H\sim 10^5\,$m, we get $\epsilon\lesssim 10^{-15}\,$m$^2$/s$^3$; and for Enceladus, with  $g=0.1\,$m/s$^2$, $S\lesssim 30\,$g/kg,  and $H\gtrsim 1\e4\,$m we get $\epsilon \lesssim 3\e{-16}\,$m$^2$/s$^3$. For Snowball Earth and Europa, these upper bound estimate are at least two orders of magnitude lower than our estimates for the energy input from vertical heatflux (table 2), suggesting that molecular diffusion even in the presence of a strong salt stratification is unlikely to be a major energy source. For Enceladus, vertical salt diffusion is also unlikely to be an important energy source although molecular diffusion may be important if the salinity is low enough for the thermal expansivity to be negative and if other potential mechanical energy sources (such as tides or librations) are negligible.  In the presence of significant tidally-driven turbulence, the tidally-driven mixing effectively replaces the molecular diffusivity, and can become a dominant energy source in all oceans. This will be discussed in the next section.  

\begin{figure}
\noindent\centering\includegraphics[width=\textwidth]{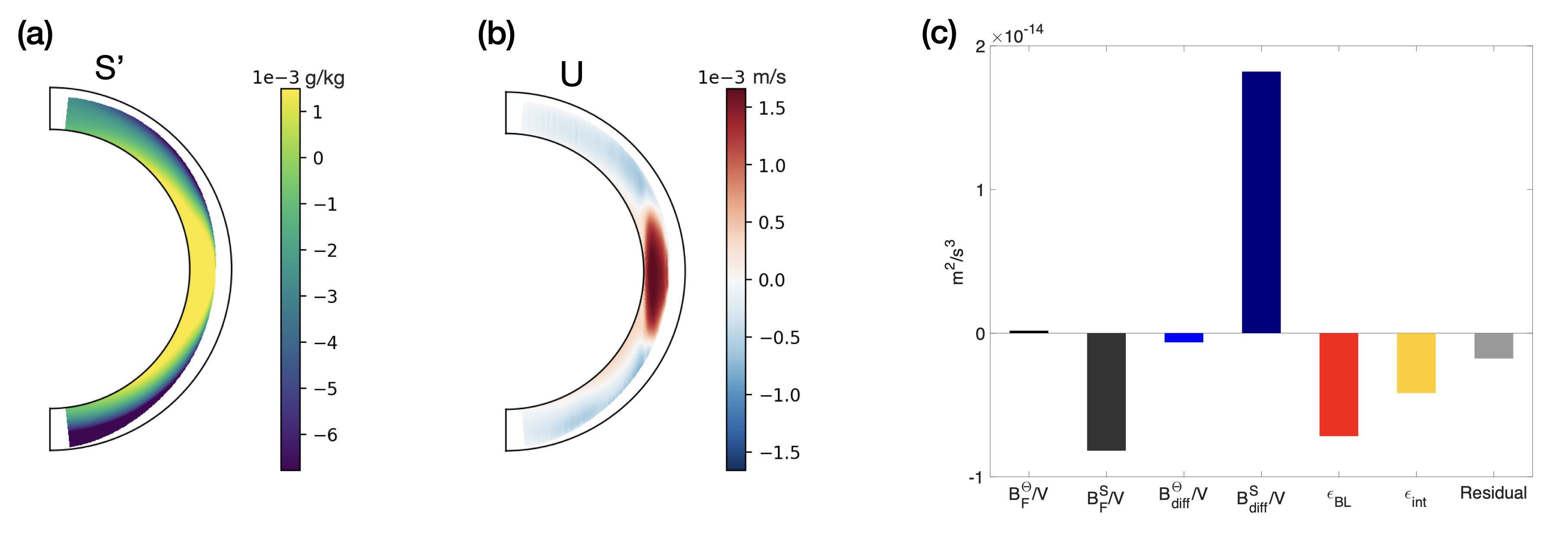}
\caption{Results for a ``salt-driven" Enceladus ocean simulation from \cite{Kang2021}.  Salt flux forcing from freezing and melting is prescribed at the bottom of an ice sheet that is thicker at the equator and thinner near the poles. Mean ocean salinity is 20 g/kg, leading to a very small thermal expansivity and the simulation does not include heating at the ocean floor. As a result, density gradients and energetics are dominated by salinity and salt fluxes. \textbf{(a)} Snapshot of Salinity and \textbf{(b)} zonal velocity as a function of depth and latitude. \textbf{(c)} Energy Budget.  From left to right are the energy sources and sinks (per unit volume) due to heat flux forcing at the boundaries ($\B^\Theta_{\mathcal{F}}/V$), salt flux forcing at the boundaries ($\B^S_{\mathcal{F}}/V$), vertical diffusion of heat ($\B^\Theta_\text{diff}/V$), vertical diffusion of salt ($\B^S_\text{diff}/V$), boundary layer dissipation ($\epsilon_\text{BL}$) and interior viscous dissipation ($\epsilon_\text{int}$). The last bar shows the residual associated with non-equilibrium effects and errors in our offline calculation of the energy budgets.}
\label{Enceladus_Salt}
\end{figure}

As for virtually all numerical simulations, the models of \cite{Zhu2017}, \cite{Kang2020,Kang2021}, \cite{AshkenazyTziperman2020} and \cite{Lobo2021} employ vertical diffusivities that are multiple orders of magnitude larger than the molecular value. In this case vertical diffusion acting against a statically stable stratification can drive a substantial circulation. This is illustrated in Fig. \ref{Enceladus_Salt} for a ``salt-driven" Enceladus Ocean simulation from \cite{Kang2021}\footnote{results shown here are from the simulation with 3D geometry and intermediate salinity $S=20$g/kg in \cite{Kang2021}}, which uses a vertical diffusivity of $\kappa=5\e{-3}\,$m$^2$s$^{-1}$. The primary source of mechanical energy in the simulations is associated with vertical diffusion of salt, which balances the energy sink associated with salt fluxes at the ice-ocean interface as well as the kinetic energy dissipation at the boundaries and in the ocean interior.  However, as in the low-salinity Enceladus simulations of \cite{ZengJansen2021} discussed above, the large vertical diffusivities need to be interpreted as representing mixing by unresolved small-scale turbulence. In reality, this turbulent mixing requires a source of kinetic energy, and the mechanistic basis for this energy source must be established in order for the results of the simulations to be sound. In Earth's ocean today the source of energy for small-scale turbulent kinetic energy ultimately comes from winds and tides \citep[e.g.][]{WunschFerrari2004}. Surface wind stress is absent in globally ice covered oceans, which leaves tides or librations as the most obvious source of turbulent kinetic energy in an ocean where buoyancy forcing is primarily due to salt fluxes at the ice-ocean interface.


\section{Tides and Turbulent Mixing} \label{sec:tides}

\subsection{Theory}

Tidal forcing generates kinetic energy in the ocean (via the second term on the LHS of Eq. \ref{global_KE}). The tidal perturbation potential is typically approximately vertically constant throughout the depth of the ocean, thus generating barotropic (i.e. vertically constant) tides. In Earth's ocean these barotropic tidal waves are largely linear until they encounter shallow shelfs, where the tidal amplitude increases and most of the tidal energy dissipation is assumed to occur \citep[][]{WunschFerrari2004}. The icy moons do not have shelf seas similar to Earth's ocean, although tidal dissipation may nevertheless be significantly enhanced in specific regions \citep[e.g.][]{HayMatsuyama2019}.  Linear tides in the deep ocean do not contribute significantly to the transport of heat, salt and other properties in the ocean, such that global models of Earth's ocean circulation can reproduce realistic large-scale circulations and tracer distributions without explicitly accounting for tides. However, it is possible that  tidal waves on some icy moons become sufficiently non-linear to lead to substantial rectified mean flows \citep[e.g.][]{Huthnance1981,Brink2011}. Since the current manuscript is focussed on buoyancy-driven circulation, we will not further consider the possible role of rectified tidal flows here, but note that this remains a potentially important topic for further research.

Despite weak direct interactions with the large-scale flow, ocean tides can have an important effect on the large-scale circulation in the presence of a statically stable stratification (e.g. driven by freshwater fluxes at the ice-ocean interface). 
Barotropic tides interacting with a rough sea-floor topography in the presence of a statically stable stratification can transfer their energy into baroclinic tides, i.e. internal waves that propagate both horizontally and vertically and can be associated with substantial vertical shears  \citep[e.g.][and references therein]{Mackinnon2017}. These shears in turn can lead to instabilities and eventually the generation of turbulence within the water column. This pathway is believed to be one of the main drivers of 3D turbulence in Earth's deep ocean \citep[][]{Vic2019}. Although most of the the turbulent kinetic energy is dissipated into heat, some fraction, usually denoted by the "mixing efficiency", $\Gamma$, is converted to potential energy via vertical mixing of the stratified water column, i.e.:
\eq
\B^t = \int  (\bar{w'b'})^{t} dV=  -\int \Gamma \epsilon^t dV ,
\qe
where $-\mathcal{B}^t$ is the potential energy generation due to tidally-driven turbulent mixing, $(\bar{w'b'})^{t}$ is the vertical buoyancy flux associated with tidally-driven turbulence, $\epsilon^{t}$ is the tidally generated turbulent kinetic energy dissipation rate per unit mass, and $\Gamma\lesssim 0.2$ is the mixing efficiency \citep[e.g.][]{PeltierCaulfield2003}.

The potential energy generated by turbulent mixing in a stratified ocean can then drive a large-scale circulation that converts the potential energy back to kinetic energy via a positive conversion $\bar{w}\bar{b}$ (where the overbars denote the large-scale circulation).  If turbulent mixing is the dominant source of potential energy (i.e. when buoyancy gain and buoyancy loss from external forcing occur at approximately the same depth) the conversion $\bar{w}\bar{b}$ is, on average, approximately equal and opposite to the downward buoyancy flux associated with tidally-driven turbulence:
\eq
\B \equiv  \int wb dV = \int \bar{w}\bar{b} dV + \int (\bar{w'b'})^{t} dV \equiv \bar{\B} + \B^t  \approx 0 .
\qe 
where $\bar{\B}$ is the kinetic energy generated (and ultimately dissipated) by the mean flow. The energy cycle can then be summarized as follows. Tides generate turbulent kinetic energy, a fraction of which is converted into large-scale potential energy via downward turbulent buoyancy flux (with the remainder being dissipated into heat).  This potential energy can then be converted back to kinetic energy (and ultimately be dissipated) by the large-scale circulation. This mechanism requires a statically stable stratification such that turbulent mixing transports buoyancy downwards. In the absence of a statically stable stratification, we still expect tidally-driven turbulence to contribute to the mixing of properties, but we do not expect it to contribute to driving a large-scale circulation. 

In addition to tides, librations and injection of fluid through fissures in the ice or solid core may provide sources of kinetic energy. Librations are expected to play a similar role to tides, and as for tides, their effect on the large-scale circulation is expected to depend on their ability to generate turbulence in the ocean interior and on the presence of a statically stable stratification. Injection of water through fissures in the boundaries (which is not formally included in the global kinetic energy-budget in Eq. (\ref{global_KE}), where we assumed no-normal-flow boundary conditions) is likely to drive mechanical turbulence primarily in the vicinity of the boundaries, but strong jets emanating from the ice shell, as proposed by \cite{KiteRubin2016}, may play an important role in the dynamics of the upper ocean\footnote{For steady planar turbulent jets in an unstratified fluid, the peak mean flow speed has been found to decay with distance from the orifice as $U_{max}\approx 2.5 \sqrt{d/x}$, where $d$ is the width of the slot from which the jet emanates and $x$ is the distance  \citep[e.g.][]{GutmarkWygnansk1976}. Turbulent flow speeds are about $20-25\%$ of this value. \cite{KiteRubin2016} suggest that Enceladus' ``tiger stripes" are associated with O($1\,$m) wide slots in which tidal forces generate O($1\,$m/s)  jets. Assuming a neutral stratification, the results for planar turbulent jets would then suggest peak mean flows on the order of $10\,$cm/s and associated turbulent velocities on the order of a few cm/s to persist to a depth of a few hundred meters into the ocean. The relatively weak depth dependence ($U_{max}\propto x^{-1/2}$) moreover indicates that weaker but significant flows may be penetrating much deeper, although results for steady jets are likely to become less applicable to the oscillatory jets proposed by \cite{KiteRubin2016} at greater depth. Stratification may further limit the penetration depth of turbulent jets.}. A full investigation of the effects of such jets is beyond the scope of this study, but remains as an interesting subject for future research.

\subsection{Application to Snowball Earth, Europa, and Enceladus}

We begin by estimating the potential role of tidally-driven turbulence in the Snowball Earth ocean.
Although little is known specifically about tidal energy dissipation during Snowball Earth periods, average tidal dissipation since the Precambrian was likely somewhat smaller but of similar order of magnitude as the present-day value \citep{Williams2000,Green2017}. The direct effect of the Snowball Earth ice sheet on tides was likely small \citep{Wunsch2016}. For scaling purposes we therefore here assume a tidal energy input on the order of 10$^{12}\,$W, which is comparable to the present day value of around 3.5$\times 10^{12}\,$W \citep{WunschFerrari2004}.  Assuming a mean ocean depth of about $2\,$km, an ocean area of 3.5$\times 10^{14}\,$m$^2$ and density $\rho_0\approx 1000\,$kg$\,$m$^{-3}$, the average tidal energy dissipation per unit mass is around $\epsilon^t\approx 10^{-9}\,$m$^2\,$s$^{-3}$, which is a little over an order of magnitude larger than the estimated energy input due to thermal forcing (cf. Table \ref{tab:para}). If a significant fraction of the dissipated tidal energy is dissipated in the stratified ocean interior and is relatively efficiently converted to potential energy (i.e. $\Gamma \gtrsim$ 10\%) it thus may play a significant role in driving a large-scale ocean circulation. Specifically, one may envision a scenario where freezing and melting at the ice-ocean interface combined with tidally-driven turbulent mixing, which pushes the light fresh water into the ocean interior, can drive a significantly stronger large-scale ocean circulation then the thermal forcing alone. This scenario appears to be relevant for the interpretation of simulation results by \cite{Ashkenazy2013} and \cite{Jansen2016}, although the turbulent vertical mixing in these simulations is parameterized with prescribed vertical diffusivities (i.e. $(\bar{w'b'})_{t}\rightarrow -\kappa \partial_z\bar{b}$) and is thus not constrained by tidal energy dissipation. The present analysis nevertheless suggests that the vertical mixing may be justifiable on energetic grounds given the expected tidally-driven turbulence, although it would require that a significant fraction of the tidal energy is dissipated in the stratified ocean interior (as opposed to being highly localized in boundary layers where the stratification is vanishingly small) and is relatively efficiently converted to potential energy.

Tidal energy dissipation on Europa has been estimated from modeling to be on the order of 10$^9$-10$^{11}\,$W \citep{Chen2014,Matsuyama2018,HayMatsuyama2019}. Assuming, as before, an ocean depth on the order of 10$^5\,$m, an area of around 3$\e{13}\,$m$^2$, and a density $\rho_0\approx 1000\,$kg$\,$m$^{-3}$, we estimate a tidal energy dissipation per unit volume of around $\epsilon^t\approx 3\e{-13}-3\e{-11}\,$m$^2\,$s$^{-3}$, which is between about one order of magnitude smaller to two orders of magnitude larger than the energy input by buoyancy forcing (cf. Table \ref{tab:para}). Depending on the estimate for tidal energy dissipation, and the fraction of the dissipated tidal energy that is converted to potential energy, tidally-driven turbulent mixing thus may or may not play a signifcant role in driving a large-scale ocean circulation on Europa. Specifically, freezing and melting at the ice-ocean interface combined with tidally-driven turbulent mixing may drive a significantly stronger large-scale ocean circulation then the thermal forcing alone, if, and only if, ocean tidal dissipation is near the upper end of these estimates and a large fraction of that energy contributes to vertical mixing against a stable stratification. A significant salt-driven circulation was found in the simulations of \cite{AshkenazyTziperman2020}, although turbulent vertical mixing is parameterized with a prescribed Earth-like vertical diffusivity in the simulations and is not constrained by tidal energy dissipation. Better estimates of tidal energy dissipation in Europa's ocean are needed to determine whether a significant turbulent vertical diffusivity can be justified for stratified regions of Europa's ocean. 

For Enceladus' ocean, energy dissipation associated with the eccentricity and obliquity tides has been estimated to be relatively small, with a total dissipation between 10-10$^4\,$W suggested by \cite{Matsuyama2018}, and \cite{HayMatsuyama2019}.  More recently, \cite{Rovira2023} argued that much larger tidal dissipation may occur in the presence of strong stratification and baroclinic tide resonances, although dissipation rates larger than 10$^6\,$W remain unlikely for the expected range of stratifications. \cite{Tyler2020} has argued that much larger dissipation rates are possible, accounting for virtually all of the observed O(10$^{10}\,$W) heat flux, although the calculations generally predict the total dissipation in the ice-ocean system, and it is likely that most of the dissipation in the suggested scenarios would indeed occur in the ice sheet \cite[c.f.][]{Beuthe2016}. Assuming that the most likely range for ocean tidal dissipation on Enceladus is $\sim$10-10$^6\,$W, and using an average ocean depth of around 3$\e{4}\,$m and an area of around $5\e{11}\,$m$^2$, we find an average tidal energy dissipation rate anywhere between around $\epsilon^t\approx 7\e{-19}-7\e{-14}\,$m$^2\,$s$^{-3}$. The upper end of this range remains more than an order of magnitude smaller than the estimated maximum mechanical energy input via thermal forcing (cf. Table \ref{tab:para}). 

However, Enceladus ocean may be more strongly affected by librations of the ice shell, which is decoupled from the solid interior \citep{Thomas2016}. It is not clear whether librations lead to relatively modest dissipation confined to the ocean-ice boundary layer, thus not contributing significantly to interior mixing, or whether the excitation of internal waves or elliptical instability (an instability that can break up elliptical streamlines) can lead to relatively strong mixing throughout the depth of the ocean \citep{Lemasquerier2017,WilsonKerswell2018,Rekier2019,Soderlund2020}. At the extreme end, \cite{WilsonKerswell2018} argue that it is possible that all of the O(10$\,$GW) of heating on Enceladus may be a result of librational dissipation, which, divided by the mass of Enceladus' ocean would give $\epsilon^t\sim 10^{-9}$m$^2\,$s$^{-3}$---many orders of magnitude larger than the possible thermal energy input. In this case it is likely that libration-driven turbulence would play the dominant role in ocean mixing. Narrowing down the large uncertainty in the energy dissipation rate associated with libration-driven turbulence will be key to constraining the turbulent diffusivity, which is an essential parameter in numerical simulations of Enceladus' ocean \citep{Kang2021,ZengJansen2021}.

\section{Conclusions} \label{conclusions}

Consideration of the sources and dissipation rates of kinetic energy provides useful constraints for the circulation of ice-covered oceans. In general, kinetic energy can be generated by tides or by conversion from potential energy, which in turn can be generated by heat and salt flux forcing. We here assumed an ocean in a statistically steady state in the sense that the mean tendencies of heat, salt and kinetic energy are small. Relaxing this assumption - e.g. by allowing for an actively growing ice sheet where the heat loss through the ice sheet is much larger than the heating from the core \citep[as proposed by][]{RobertsNimmo2008} would require the inclusion of tendency terms in our budgets, which in turn would lead to a significantly less well constrained problem. 

A commonly assumed forcing consists of heating from the solid core balanced by heat loss through the ice sheet, which acts as a source of potential energy and can drive an ocean circulation, as long as the thermal expansivity of the water is positive. In the oceans of Snowball Earth, Europa, and Enceladus, the associated energy input, however, is orders of magnitude smaller than the wind energy input that dominantly drives the circulation in Earth's present-day ocean.  We predict that the resulting thermally-driven flows have flow speeds of at most a few cm/s and will be strongly affected by rotation (i.e small Rossby numbers). Numerical simulations of thermally-driven flows on icy moons, which typically use artificially large viscous dissipation and sometimes artificially large thermal forcing may misrepresent both energy sources and sinks by multiple orders of magnitude, which can lead to widely different and unrealistic levels of kinetic energy. 

Salt fluxes associated with freezing and melting at the ice sheet boundary only provides a source of potential energy if freezing occurs under thinner ice than melting, which is unlikely in equilibrium where freezing and melting needs to be in balance with ice flow. In the more likely scenario where melting on average occurs at a shallower depth than freezing, the salt flux forcing acts as a sink of potential energy. Vertical mixing, which can push the lighter fresh water into the interior, is then needed to drive a  circulation. Molecular diffusion can in theory accomplish such mixing, but the associated energy source is likely to be relatively small. Much stronger turbulent mixing may be driven by tidal energy dissipation.

Ocean tides and librations may provide a key energy source for ocean turbulence. Current estimates indicate that tidally-driven vertical mixing is likely to be important in a Snowball Earth ocean, and could possibly play a significant role on Europa, while librations may provide a key source of turbulent kinetic energy on Enceladus. However, the magnitude and spatial distribution of turbulence generated by tides and librations remains highly uncertain, which represents a major hurdle to better constrain the circulations of icy-world oceans. An improved understanding of ocean tides and librations on icy moons thus remains as an important topic for future research.

\appendix

\section{A generalized expression for energy input from buoyancy forcing} \label{AppA}

To quantify the role of heat and salinity forcing on the vertical buoyancy flux, while fully accounting for nonlinearities in the equation of state, it is useful to introduce the dynamic enthalpy \citep{Young2010}:
\eq
h^{\ddagger}(\Theta,S,z)=\int_z^{0} \tilde{b}(\Theta,S,z')dz' . 
\qe
The dynamic enthalpy, $h^\ddagger$, is closely related to the potential energy, and indeed reduces to the well known expression for the potential energy in the limit of an equation of state with no explicit depth dependence (i.e. $b=\tilde{b}(\Theta,S)$) where $h^{\ddagger}\rightarrow b z$. 


The evolution equation for $h^\ddagger$ is
\eq
\frac{Dh^\ddagger}{Dt}= -wb + \dot{\Theta}\partial_\Theta h^{\ddagger} + \dot{S} \partial_S h^{\ddagger} ,
\qe
where $\dot{\Theta}\equiv D\theta/Dt$ and $\dot{S}\equiv DS/Dt$ are given by Eqs. (\ref{DTdt}) and (\ref{DSdt}).

Integrating globally and assuming the global enthalpy budget to be in equilibrium, we can relate the globally integrated vertical advective buoyancy flux, $\B$, which provides a source of KE, to the thermal and salinitiy forcing, $\dot{\Theta}$ and $\dot{S}$:  
\eq
\B \equiv  \int wb dV = \int \left(\dot{\Theta}\partial_\Theta h^{\ddagger}+ \dot{S}\partial_S h^{\ddagger} \right) dV .
\label{B_irrev1}
\qe
Defining
\eq
\widehat{g\alpha}(\Theta,S,z)\equiv  \frac{1}{-z} \partial_\Theta h^{\ddagger} = \frac{1}{-z}\int_z^{0}g \alpha(\Theta,S,z')dz'  
\qe
and 
\eq
\widehat{g\beta}(\Theta,S,z)\equiv  \frac{1}{-z} \partial_S h^{\ddagger} = \frac{1}{-z}\int_z^{0}g \beta(\Theta,S,z')dz'  ,
\qe
we can write Eq. (\ref{B_irrev1}) as
\begin{eqnarray} \nonumber
\B  &=& -\int \left( \dot{\Theta}\widehat{g\alpha} z + \dot{S}\widehat{g\beta}z  \right) dV \\  \nonumber
&=&- \int \kappa \left( \nabla^2\Theta\,\widehat{g\alpha} z + \nabla^2S \,\widehat{g\beta}z  \right) dV \\  \nonumber
&=&- \int \kappa \left( \nabla\cdot  (\nabla \Theta\,\widehat{g\alpha} z) - \nabla \Theta \cdot \nabla(\widehat{g\alpha} z)  \nonumber
                                + \nabla\cdot(\nabla S \,\widehat{g\beta}z)  - \nabla S\cdot\nabla (\widehat{g\beta} z)  \right) dV \\     \nonumber 
 &=&  \int \frac{\mathcal{Q}_{top}}{\rho c_p} \widehat{g\alpha} z_{top} dA -  \int \frac{\mathcal{Q}_{bot}}{\rho c_p} \widehat{g\alpha} z_{bot} dA \\   \nonumber
 & &       + \int \mathcal{S}_{top}\widehat{g\beta} z_{top} dA  -  \int \mathcal{S}_{bot} \widehat{g\beta} z_{bot} dA \\  \nonumber 
 & &+ \int g \kappa \left( \alpha  \partial_{z}  \Theta+ \beta  \partial_{z}S \right) dV  \\                     
 & &+ \int z\kappa \left( \nabla_h (\widehat{g\alpha}) \cdot \nabla_h \Theta + \nabla_h(\widehat{g\beta}) \cdot \nabla_h S  \right) dV  .   
\label{B_irrev2}                    
\end{eqnarray}
Eq. (\ref{B_irrev2}) allows us to relate the globally integrated advective upward buoyancy flux (and hence the associated KE generation) to the heat and salt fluxes through the boundaries (the first four terms on the RHS of Eq. \ref{B_irrev2}), reduced by the diffusive buoyancy flux (the fifth term on the RHS of Eq. \ref{B_irrev2}).   The last term in Eq. (\ref{B_irrev2}) captures the effect of buoyancy sources or sinks that arise from horizontal diffusive mixing of $\Theta$ and $S$ due to the nonlinearity in the equation of state (the cabbeling effect). 

The potential role of the cabbeling term in Eq. (\ref{B_irrev2}) can be illustrated by considering again the example of section \ref{sec:low_salt}, which may resemble the conditions on Enceladus, assuming a relatively low salinity for the latter. That is we assume a stratified layer with negative thermal expansion coefficient above a convective deep layer with weakly positive thermal expansion coefficient. For illustrative purposes, let us assume the upper and lower boundaries to be horizontal, bottom and surface heat fluxes to be spatially uniform, and we ignore radial variations in the surface area, such that $Q_{top}=Q_{bot}\equiv Q$, as well as vertical variations in gravity. Since variations in the thermal expansion coefficient are primarily due to differences in the temperature, we moreover neglect the direct depth dependence of $\tilde{b}$ and assume that variations in salinity are small, such that $\widehat{g \alpha}=g \alpha(\theta)$. In this case, Eq. (\ref{B_irrev2}) simplifies to
\begin{eqnarray} \nonumber
\B  &=& \frac{gQ}{\rho c_p} \int \left(\alpha(\Theta_{top}) z_{top} -\alpha(\Theta_{bot}) z_{bot}\right) dA \\   \nonumber
 & &+ \int g \kappa \alpha(\Theta) \partial_{z}  \Theta  dV + \int g z \kappa \partial_\Theta{\tilde{\alpha}} |\nabla_h \Theta|^2 dV   \,.                        
\label{B_irrev_enc}
\end{eqnarray}
where we used that $\nabla_h \alpha \cdot \nabla_h \Theta = \partial_\Theta{\alpha}  |\nabla_h \Theta|^2$. The last term can now be associated with the diffusive destruction of temperature variance, which leads to a buoyancy source if $\partial_\Theta{\alpha} >0$ (amounting to a positive curvature in  $b(\theta)$).

Notice that the magnitude (and potentially the sign) of the first and last term in Eq. (\ref{B_irrev_enc}) depend of the choice of reference level. E.g. setting $z_{bot}=0$ and $z_{top}=H$ where $H$ is the depth of the ocean (which we assume to be constant with flat upper and lower boundaries) we find   
\begin{eqnarray} \nonumber
\B  &=& \frac{gQH}{\rho c_p}   \int \alpha(\Theta_{top}) dA \\  
 & &+ \int g\kappa \alpha(\Theta) \partial_{z}  \Theta  dV + \int_0^H gz \kappa \partial_\Theta{\alpha} |\nabla \Theta|^2 dV   \,.    
 \label{B1_enc}                    
\end{eqnarray}
Defining the reference level instead such that $z_{top}=0$ and $z_{bot}=-H$ we get 
\begin{eqnarray} \nonumber
\B  &=&  \frac{gQH}{\rho c_p}  \int \alpha(\Theta_{bot}) dA \\   
 & &+\int g \kappa \alpha(\Theta) \partial_{z}  \Theta  dV + \int_{-H}^0  gz \kappa \partial_\Theta{\alpha} |\nabla \Theta|^2dV   \,.                        
 \label{B2_enc}                    
\end{eqnarray}
The first term differs dramatically in the two equations, as $\alpha(\Theta_{top})<0$ while $\alpha(\Theta_{bot})>0$. 
This difference is compensated by the last term, which also changes sign, as $z>0$  in (\ref{B1_enc}) and $z<0$  in (\ref{B2_enc}). 
The last term in Eq. (\ref{B_irrev2}) hence can be of leading order importance if variations in the thermal (or haline) expansion coefficient are large. Since this term cannot be predicted based on knowledge of the boundary conditions alone, this most general formulation is hence less immediately useful as a predictor of flow energetics. However, if horizontal temperature variations are relatively small, Eq. (\ref{B2_T})  remains a useful approximation and allows us to estimate the kinetic energy input without explicit consideration of diffusive variance destruction. 
\bibliography{references}{}
\bibliographystyle{aasjournal}

\end{document}